\documentclass[aps,prc,twocolumn,letterpaper,superscriptaddress,nofootinbib,showpacs,floatfix,10pt]{revtex4-1}
\usepackage{graphicx}
\usepackage{comment}
\usepackage{setspace}
\usepackage{amsmath}
\usepackage{amssymb}
\usepackage[textwidth=16cm,textheight=22cm]{geometry}
\usepackage{color}
\usepackage{indentfirst}
\usepackage{xspace}
\usepackage{hyperref}
\usepackage{verbatim}
\usepackage{epstopdf}
\usepackage{graphicx}
\usepackage{longtable}
\usepackage{lineno}
\usepackage{subfigure}

\newcommand{\pt}   {\ensuremath{p_\mathrm{T}}\xspace}

\newcommand{\gp}{\ensuremath{\gamma\text{p}}\xspace}
\newcommand{\ampt} {\textsc{ampt}\xspace}
\newcommand{\epos} {\textsc{epos}\xspace}
\newcommand{\pythia} {\textsc{pythia8}\xspace}
\newcommand{\nch}   {\ensuremath{N_\text{ch}}\xspace}

\newcommand{\mpt}   {\ensuremath{\langle p_{\mathrm{T}} \rangle}\xspace}
\newcommand{\meanpt}   {\ensuremath{\langle p_{\mathrm{T}} \rangle}\xspace}

\begin{document}
%\linenumbers

%\preprint{}

%\title{Investigation of differential two-particle correlations through the balance functions in photon-induced processes using PYTHIA 8}

%\title{Investigation of differential two-particle correlations through the balance functions in photon-induced processe using PYTHIA 8}

\title{Exploring differential two-particle correlations in $\gamma$p and low-multiplicity pp collisions using PYTHIA8}

%Investigation of  two-particle differential charged particle correlation in pp and gammap collisions using PYTHIA8

\author{Subash~Chandra~Behera}
\email{Contact author: subash.chandra.behera@cern.ch}
\affiliation{INFN–Sezione di Roma, Piazzale Aldo Moro, 2-00185 Rome, Italy}

\author{Dukhishyam~Mallick}
\email{Contact author: dukhishyam.mallick@cern.ch}
\affiliation{INFN–Sezione di Cagliari, Monserrato, 09042, Cagliari, Italy}

%add author here--
%\author{Author Name}
%\email{email ID}
%\affiliation{Institute with Complete with address}

%\date{\today}

\begin{abstract}
A study of two-particle differential number ($B$) and transverse momentum ($P_{2}^\mathrm{CD}$) balance functions in photon-proton (\gp) and proton-proton (pp) collisions at $\sqrt{s}=$ 5.36 TeV is presented. The analysis focuses on inclusive charged hadrons within the pseudorapidity coverage $|\eta|<2.4$ and the transverse momentum interval $0.3 < p_\mathrm{T} < 3.0$ GeV and examines their correlations in terms of relative pseudorapidity ($\Delta\eta$) and relative azimuthal angle ($\Delta\phi$). The correlation functions are evaluated for same- and opposite-sign pairs, and their combinations are used to extract charge-dependent (CD) and charge-independent (CI) components. The evolution of the nearside peak of the CD correlations is investigated in terms of $\Delta\eta$ and $\Delta\phi$ as a function of charged-particle multiplicity (\nch) for \gp collisions and compared to pp collisions at a similar multiplicity range. A clear multiplicity dependence of the balance-function width is obtained. The width is found systematically lower in \gp events than in pp collisions. This study provides valuable information on particle correlations and production mechanisms in low-\nch regimes for upcoming measurements in small systems.

\end{abstract}


\maketitle

\section{Introduction}
Measurements from ultrarelativistic heavy-ion collisions at the Relativistic Heavy Ion Collider (RHIC) and the Large Hadron Collider (LHC) have provided compelling evidence for the formation of a strongly interacting matter known as the quark-gluon plasma (QGP)~\cite{Harris:2023tti, ALICE:2022wpn, CMS:2024krd, QGPmedium1, starqgp, ALICE:2020fuk, qgpmed2}.  \\
%The observation of pronounced collective flow, especially the large elliptic flow coefficient ($v_2$), has been observed.
%The observation of  collective flow, especially the large elliptic flow coefficient ($v_2$), has been measured and comparing calculation from theoretical descriptions based on relativistic hydrodynamics, it is found that the QGP behaves like an almost perfect fluid with extremely low shear viscosity

The observation of collective flow, especially the large elliptic flow coefficient ($v_{2}$), and its comparison with theoretical calculations based on relativistic hydrodynamics suggests that the QGP behaves like an almost perfect fluid with extremely low shear viscosity. In particular, the large value of $v_{2}$ is found to be consistent with a medium having a very small shear viscosity to entropy density ratio, $4\pi\eta/s \sim 1$--$2$, close to the conjectured lower bound for a strongly coupled fluid~\cite{Kovtun:2004de,Song:2010mg}.
Studying correlations between oppositely charged particle pairs created in heavy-ion collisions provides insight into the dynamics of the QGP and the underlying hadronization mechanism~\cite{ALICE:2015nuz, scottpratt, BassScott, Pratt2012dz, jetBF, alicep2r2pp, alicebfpbpbpb, alicethree, STAR:2010plm}. The particle production is governed by local charge conservation, in which the creation of a positively charged particle must be accompanied by the production of a negatively charged partner. The balance function is designed to quantify the correlations in phase space variables such as pseudorapidity difference ($\Delta\eta$) and azimuthal angle difference ($\Delta\phi$)~\cite{cmsbf, alicebfpbpbpb, aliceoldpbpb, scottpratt, BassScott, Pratt2012dz, jetBF, Pruneau:new, Pruneau:2019baa, ALICE:pruneau}. \\

Measurements of the nearside peak width of the charge balance function exhibit a strong system-size dependence across various collision systems spanning over RHIC to LHC energies~\cite{aliceoldpbpb, cmsbf, alicethree, STAR:2010plm}. A significant narrowing of the balance functions width is observed in central heavy-ion collisions. This effect is interpreted as the signature of delayed hadronization in the presence of the QGP medium, where balancing charges are produced at a later stage of the system evolution and therefore remain closer in phase space~\cite{aliceoldpbpb, cmsbf, alicethree, STAR:2010plm}.\\

Recent observations of long-range rapidity correlations (“ridgelike” structures), nonzero flow harmonics, and the narrowing of the nearside width of the correlation functions in small collision systems exhibit features qualitatively similar to those seen in heavy-ion collisions~\cite{CMS:2024krd, CMSpp_longrange, qgpcms_2pc, Belle:2022fvl, gqp_hin11001, cmshin12015}. The unexpected narrowing behavior challenges the conventional interpretation associated with delayed hadronization in the deconfined medium. As a result, the mechanism responsible for this behavior in small systems remains not fully understood~\cite{alice_lowmultpp, qgpcms_2pc, cmshin12015}.\\

Several theoretical studies using event generators such as \textsc{pythia}, \ampt, and \epos, which incorporate distinct physics mechanisms including string fragmentation, partonic transport, and hydrodynamic evolution also reproduce the similar decreasing trend of the balance-function width with increasing multiplicity~\cite{Behera:2025ymi, jetBF, evtshape, Sahoo:2018uhb, cmsbf}. This suggests that multiple mechanisms, such as local charge conservation, string dynamics, multiple parton interactions (MPIs), color reconnection (CR), or parton transport effects, may contribute to the observed behavior. \\

In this work, the correlation functions are further investigated in even smaller collision systems, extending to photon-induced \gp collisions using \pythia simulations~\cite{pythia_ref, cmspythia}. Such collisions provide a cleaner environment compared to hadronic pp interactions, as electromagnetic processes dominate and the underlying hadronic event activity is significantly suppressed~\cite{CMS2pcgp, ATLAS:2pc}. These events are characterized by very low charged-particle multiplicity due to the suppression of the MPI and the absence of an extended QGP-like medium. Therefore, \gp collisions offer an ideal baseline for studying correlation functions and disentangling genuine charge conservation effects from medium-induced phenomena observed in hadronic collisions.\\

In this work, both charge-independent (CI) and charge-dependent (CD) two-particle number and transverse-momentum (\pt) correlations are presented. The focus of this study is the CD correlation, which isolates correlations associated with charge conservation by subtracting same-sign (SS) pair contributions from opposite-sign (OS) pairs. Then, to quantify the correlation strength, the root-mean-square (RMS) widths of the CD correlation functions are extracted from their one-dimensional distributions onto $\Delta\eta$ and $\Delta\phi$.\\

This paper is organized as follows. Section~\ref{anamethod} describes the analysis methodology. Section~\ref{modeldes} discusses the model setup and framework used for this study. Section~\ref{resultssec} presents the results for $B$, $P_{2}^\mathrm{CD}$ correlations, and their corresponding widths in $\Delta\eta$ and $\Delta\phi$. Finally, Sec.~\ref{summ} summarizes this work. Additional discussions about the CI correlations can be found in Sec.~\ref{append}.

\section{Analysis Methodology}
\label{anamethod}
Two-particle number correlation functions $C_{2}$ are constructed using the “trigger” and “associated” tracks that are selected from the same event. Both the trigger and associated particles are selected in $|\eta| < 2.4$ and $0.3<p_\mathrm{T}<3.0$ GeV intervals. For each pair, the differences in $\Delta \eta$ and $\Delta \phi$ are calculated. The per-trigger associated yield in the same event is defined as
\begin{equation} \label{eqn_signal}
  S(\Delta \eta,\Delta \phi) = \frac{1}{N_\text{trig}}\frac{d^{2}N^\text{same}}{{d\Delta \eta} \ {d\Delta \phi}},
\end{equation}
where $N_\text{trig}$ is the number of trigger particles and $N^\text{same}$ is the number of same-event pairs. The mixed-event distributions are calculated by randomly mixing six chosen events in the similar multiplicity class for $0.3 < \pt < 3.0$ GeV and $|\eta| < 2.4$ intervals. This can be written as 
\begin{equation} \label{eqn_signal}
  M(\Delta \eta,\Delta \phi) = \frac{1}{N_\text{trig}}\frac{d^{2}N^\text{mix}}{{d\Delta \eta} \ {d\Delta \phi}},
\end{equation}
where $N^\text{mix}$ denotes the number of mixed-event pairs. The mixed-event background distribution accounts for the finite detector acceptance and pair efficiency~\cite{CMS2pcgp, cmsbf, CMSpp_longrange, gqp_hin11001, aliceoldpbpb, Behera:2025ymi, Belle:2022fvl}. The two-particle number correlation function is then obtained as the ratio of the same-event signal to the mixed-event background~\cite{CMS2pcgp, cmsbf, CMSpp_longrange, gqp_hin11001, aliceoldpbpb, Behera:2025ymi, Belle:2022fvl}. 
\begin{align} \label{eqn_2pc}
\frac{1}{N_\text{trig}}\frac{d^{2}N^\text{pair}}{{d\Delta \eta} \ {d\Delta \phi}}& =C_{2}(\Delta\eta, \Delta\phi) \\
&= M(0,0) \frac{S(\Delta\eta, \Delta\phi)}{M(\Delta\eta, \Delta\phi)},
\end{align}
where the term $M(0,0)/M(\Delta\eta, \Delta\phi)$ mainly accounts for the pair acceptance effects and $M(0,0)$ denotes the mixed-event yield when both particles are emitted approximately in the same direction, leading to the highest pair detection efficiency.\\

Similarly, the two-particle transverse momentum correlator, $P_{2}$ that takes care of \pt fluctuations can be defined as 
\begin{equation} \label{equ_p2}
\centering
P_{2} (\Delta\eta, \Delta\phi) = \frac{\langle \Delta p_\text{T} \Delta p_\text{T} \rangle(\Delta\eta, \Delta\phi)}{\langle p_\text{T} \rangle ^{2}},
\end{equation}

where $\Delta p_\text{T} = p_\mathrm{T} -\langle p_\mathrm{T} \rangle$ and $\langle \Delta p_\text{T} \Delta p_\text{T} \rangle$ is the differential correlator that can be defined as

\begin{equation} 
\label{equ_dptdpt}
\scriptsize
\langle \Delta p_T \Delta p_T \rangle (\Delta\eta, \Delta \phi) =  
\frac{\int_{\pt, min}^{\pt, max} \Delta \pt \Delta \pt \; c_{2}' (\vec{p_1}, \vec{p_2}) \; d\pt d\pt}  
{\int_{\pt, min}^{\pt, max} c_{2}' (\vec{p_1}, \vec{p_2}) \; d\pt d\pt },
\end{equation}
where $c_{2}'$ is the two-particle density distribution, expressed as a function of the transverse momenta. The momentum correlator is normalized $\langle \pt \rangle^{2}$ to make the $P_{2}$ correlation dimensionless. The $P_{2}$ correlation is influenced by the presence of number correlations and the transverse momentum distribution of the correlated particle pairs, and the details can be found in Refs.~\cite{alicebfpbpbpb, alicep2r2pp, Behera:2025ymi}. \\

The CI and CD components are constructed using $C_{2}$ and $P_{2}$ correlation functions from their corresponding SS and OS pairs. The CI correlation is the average of SS and OS correlation functions, and the CD correlation is the difference of the SS and OS correlation functions. The SS and OS pairs are influenced by different physical mechanisms; taking their combination can isolate the correlation associated with balancing charges. This effect is quantified through the CD correlation. 
Mathematically, these components are expressed as follows:

\begin{equation} \label{eqn_balfun_c2ci}
C_{2}^\mathrm{ CI} =\frac{1}{2}[C_{2}^{(+, -)}  + C_{2}^{(-, +)} + C_{2}^{(-, -)} + C_{2}^{(+, +)}],
\end{equation}
\begin{equation} \label{eqn_balfun_p2}
P_{2}^\mathrm{ CI} =\frac{1}{2}[P_{2}^{(+, -)}  + P_{2}^{(-, +)} + P_{2}^{(-, -)} + P_{2}^{(+, +)}],
\end{equation}

\begin{equation} \label{eqn_balfun_B}
B =\frac{1}{2}[C_{2}^{(+, -)}  + C_{2}^{(-, +)} - C_{2}^{(-, -)} - C_{2}^{(+, +)}],
\end{equation}
\begin{equation} \label{eqn_balfun_p2cd}
P_{2}^\mathrm{ CD} =\frac{1}{2}[P_{2}^{(+, -)}  + P_{2}^{(-, +)} - P_{2}^{(-, -)} - P_{2}^{(+, +)}].
\end{equation}
The strength of the balance functions is quantified by the RMS value of their one-dimensional (1D) distributions of $\Delta\eta$ and $\Delta\phi$. The 1D $\Delta\eta$ and $\Delta\phi$ projections are obtained from the two-dimensional correlation functions by integrating over the intervals $|\Delta\phi| < \pi/2$ and $|\Delta\eta| < 1.0$, respectively. Mathematically, the RMS width ($\sigma_{\mathcal{O}}$) is defined as follows:
\begin{equation}
\sigma_{\mathcal{O}} =
\sqrt{
\frac{
\sum_{i} \mathcal{O}_i^{2}\, B(\mathcal{O}_i)
}{
\sum_{i} B(\mathcal{O}_i)
}
},
\label{eq:BFwidth}
\end{equation}
where $B(\mathcal{O}_i)$ denotes the balance function evaluated in the $i\mathrm{th}$ bin centered at $\mathcal{O}_i$, and $\mathcal{O}$ represents the relative separation variable, such as $\Delta\eta$ or $\Delta\phi$. The summation runs over all bins within the selected kinematic range. A similar approach is followed to extract the RMS width of $P_{2}^\mathrm{CD}$ correlations. The $\sigma_{\Delta\eta}$ and $\sigma_{\Delta\phi}$ are calculated in $|\Delta\eta| < 2.0$ and $|\Delta\phi| < \pi/2$ intervals, respectively.

\section{Model description}
\label{modeldes}
In this study, events are generated using \textsc{pythia}~8.36~\cite{Bierlich:2022pfr, pythia_ref} to simulate both \gp and pp collisions, with approximately $0.2\times 10^{9}$ and $3\times 10^{9}$  events, respectively. For $\gamma$p collisions, the photon flux is modeled using the equivalent photon approximation implemented in \textsc{pythia}, where the photon is treated as another beam particle(\texttt{Beams:idA = 22}). The details of the modeling can be found in Ref.~\cite{Helenius:2024vdj}.\\

To provide a realistic description of photon-induced interactions and to suppress contributions from underlying-event activity and MPI, all hard QCD processes are disabled (\texttt{HardQCD:all = off}, \texttt{PartonLevel:MPI = off}), while photon-induced processes are enabled (\texttt{PhotonParton:all = on}). These settings are considered as the default tune for \gp collisions. In addition, the Monash tune for minimum-bias (MB) inelastic pp collisions at LHC energies is considered for comparison. The same tune is also applied to the $\gamma$p simulations to study the differences in final-state particle production and production mechanisms when changing the collision system from pp to $\gamma$p.\\

This configuration uses the following tunes:
\texttt{Tune:pp = 14}, (Monash)
\texttt{SoftQCD:inelastic = on},
\texttt{PartonLevel:MPI = on}, and
\texttt{ColourReconnection:reconnect = on}. This setup is referred to as the \textsc{mb}-like tune throughout this paper.\\

\begin{figure}[htbp]
  \centering
  \includegraphics[width=0.5\textwidth]{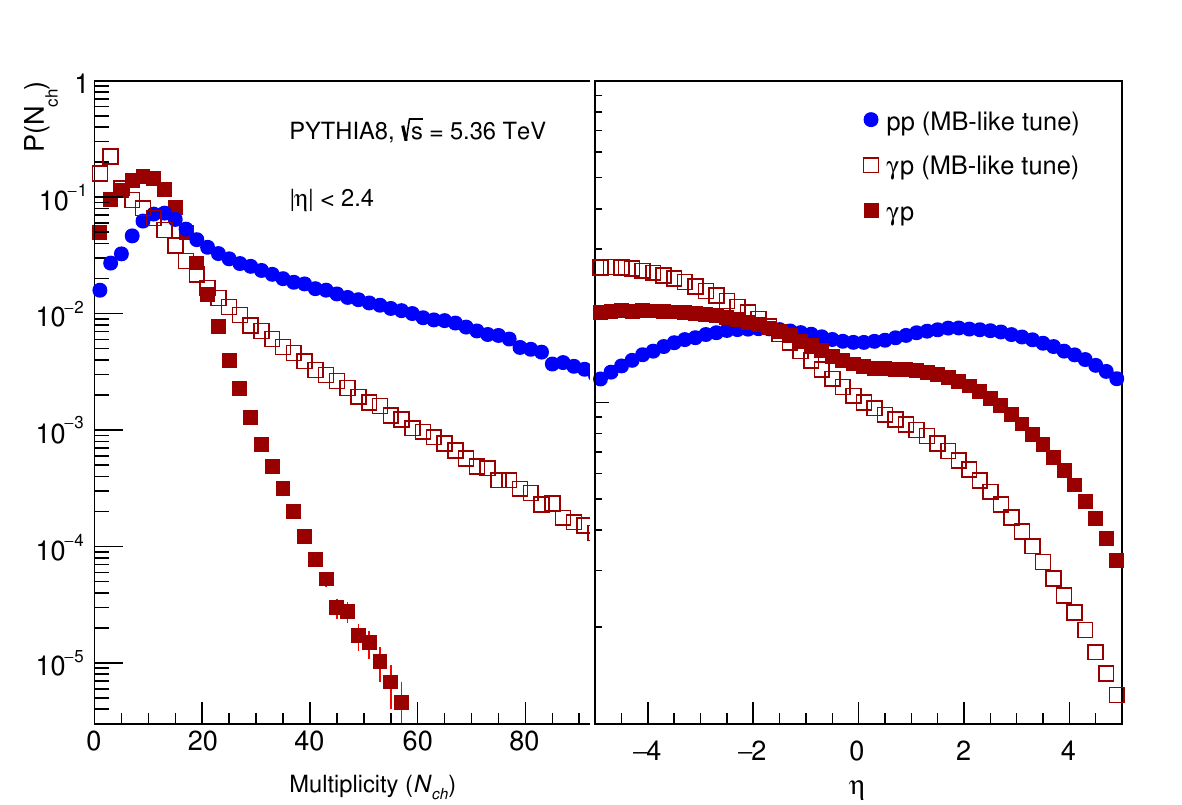}
  \caption{Comparison of multiplicity (\nch) and pseudorapidity ($\eta$) distributions of charged hadrons in MB pp, \textsc{mb}-like tune \gp, and \gp collisions at $\sqrt{s}$ = 5.36 TeV.}
  \label{fig:nch_eta}
\end{figure}

\begin{figure}[htbp]
  \centering
  \includegraphics[width=0.5\textwidth]{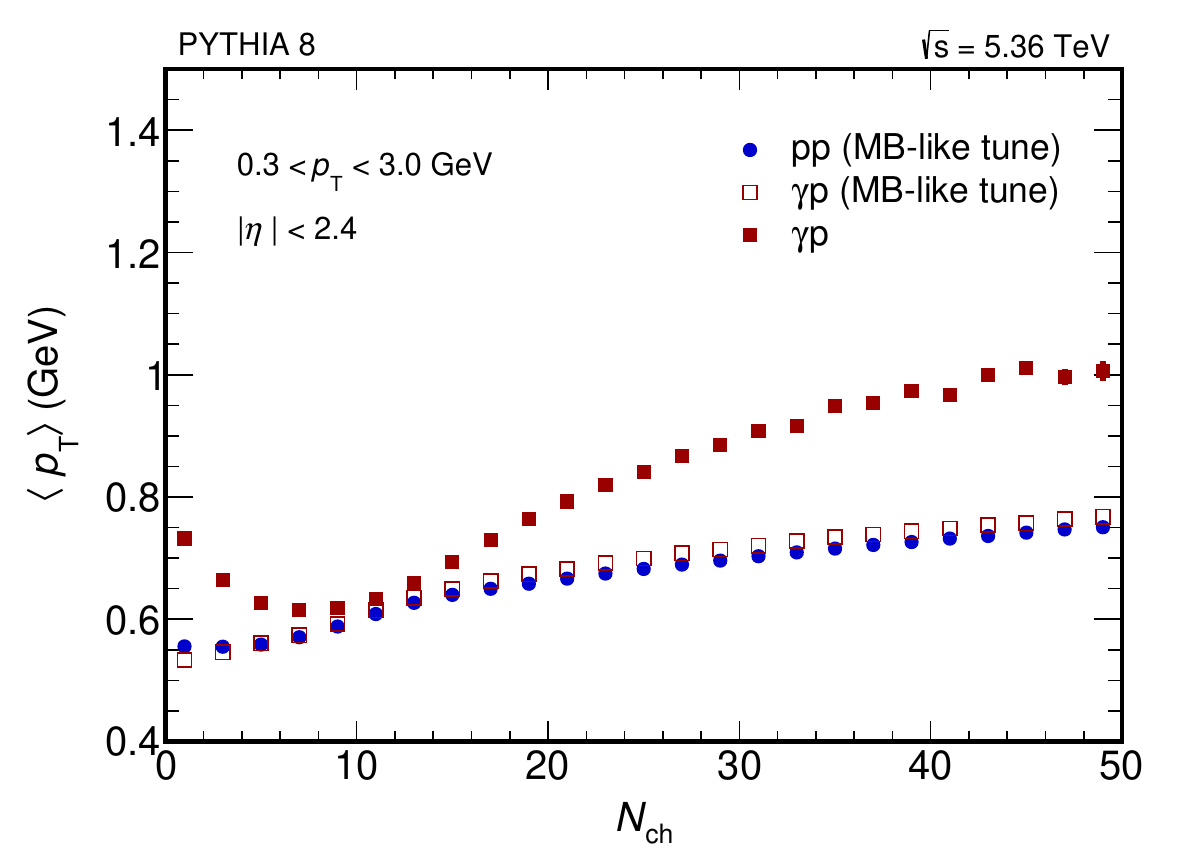}
  \caption{Comparison of average transverse momentum (\mpt)  as function of \nch for charged hadrons in MB pp, \textsc{mb}-like tune \gp and \gp collision at $\sqrt{s}$ = 5.36 TeV. Results are calculated for $|\eta|< 2.4$ and $0.3 < p_{\mathrm{T}} < 3.0$ GeV.}
  \label{fig:nch_pT}
\end{figure}

Figure~\ref{fig:nch_eta} illustrates the normalized charged-particle multiplicity (\nch, left panel) and pseudorapidity ($\eta$, right panel) distributions in pp and \gp collisions from \pythia simulations. The charged-particle multiplicity distributions for MB inelastic pp collisions show significantly higher values of \(N_{\mathrm{ch}}\) compared to \gp collisions. The distributions are evaluated within the pseudorapidity interval \(|\eta| < 2.4\). The ordering of the multiplicity spectra is observed as
\begin{equation}
N_{\mathrm{ch}}^{\gamma p} < N_{\mathrm{ch}}^{\gamma p, ~\mathrm{\textsc{mb}\text{-}like}} < N_{\mathrm{ch}}^{pp, ~\mathrm{\textsc{mb}\text{-}like}}.
\end{equation}

The MB pp collisions reach larger \nch values and exhibit broader multiplicity distribution due to the presence of multiple partonic interactions, underlying event activities, and the larger available phase space in a fully hadronic collision system. In contrast, the \gp collisions show lower multiplicity coverage and a rapidly falling distribution with increasing \nch. This behavior reflects that the particle production in photon-induced interactions is typically driven by a single hard or semihard scattering process, although the overall \nch not only depends on the colliding systems and energies but is also sensitive to the choice of tunes and underlying physics processes that are included in the simulation.\\

The $\eta$ distributions are symmetric around midrapidity ($\eta \sim$ 0) in pp collisions, whereas an asymmetric shape is observed in \gp collisions. It is expected due to the \gp configuration. The multiplicity coverage for this study is chosen based on the maximum \nch reach in \gp, and an equivalent multiplicity range is selected for pp collisions. \\

Figure~\ref{fig:nch_pT} presents the average transverse momentum (\meanpt) as a function of the \nch for \gp and pp collisions at $\sqrt{s}=5.36$~TeV from \pythia simulations. The \meanpt are calculated for charged hadrons in the interval $0.3 < p_{\mathrm{T}} < 3.0$~GeV and within $|\eta|<2.4$. A monotonically increasing behavior of \meanpt with \nch is observed.  The \meanpt values and their dependence on \nch are found to be similar between \gp and pp collisions when they have MB-like configurations. However, a steeper rise of \meanpt with \nch is observed for $\gamma$p collisions with a more realistic configuration for photon-induced interaction. The \meanpt values are higher compared to pp collisions at a given multiplicity, and the difference becomes more pronounced at higher $N_{\mathrm{ch}}$. This indicates that the \meanpt is more sensitive to the underlying physics processes and model tunes than to the nature of the colliding systems when comparing pp and \gp collisions. These results demonstrate that the correlation between multiplicity and transverse momentum depends on several components, such as the roles of quark- and gluon-initiated interactions, color reconnection, and MPI dynamics. 

\section{Results and discussion}
\label{resultssec}
\subsection{Charge-dependent correlations}
\begin{figure*}  % Wide figure spanning both columns
    \centering
    \includegraphics[width=0.3\textwidth]{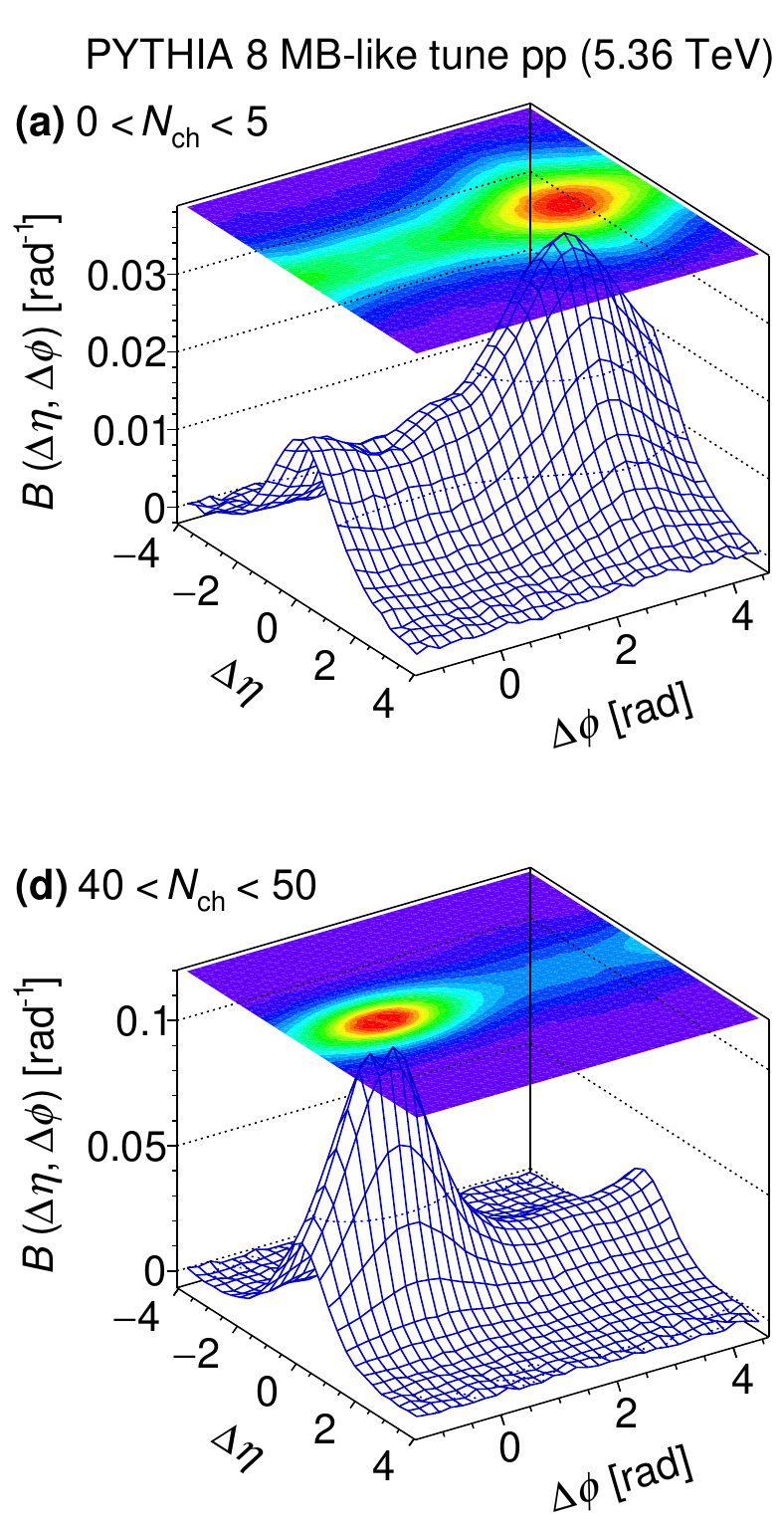}
    \includegraphics[width=0.3\textwidth]{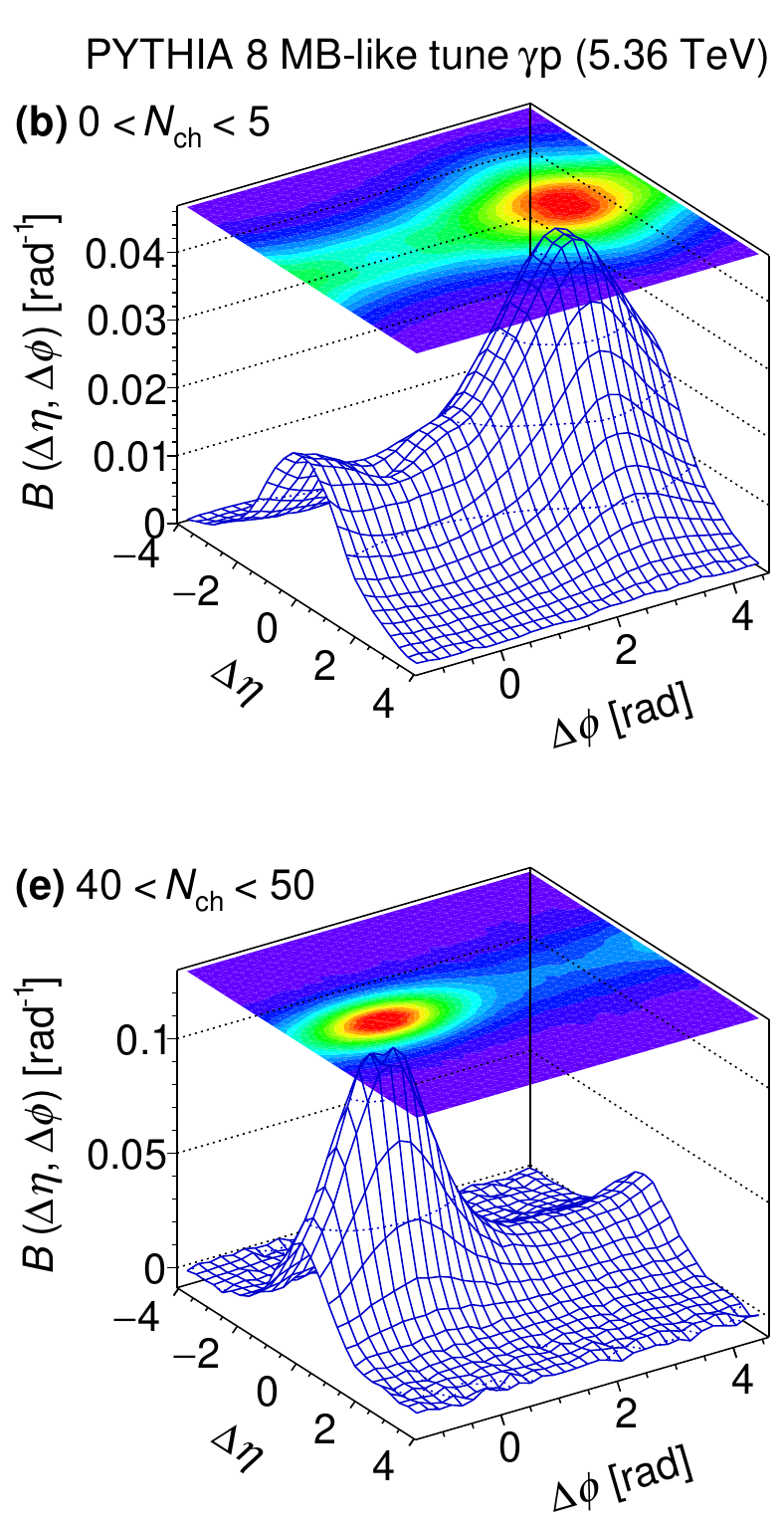}
    \includegraphics[width=0.3\textwidth]{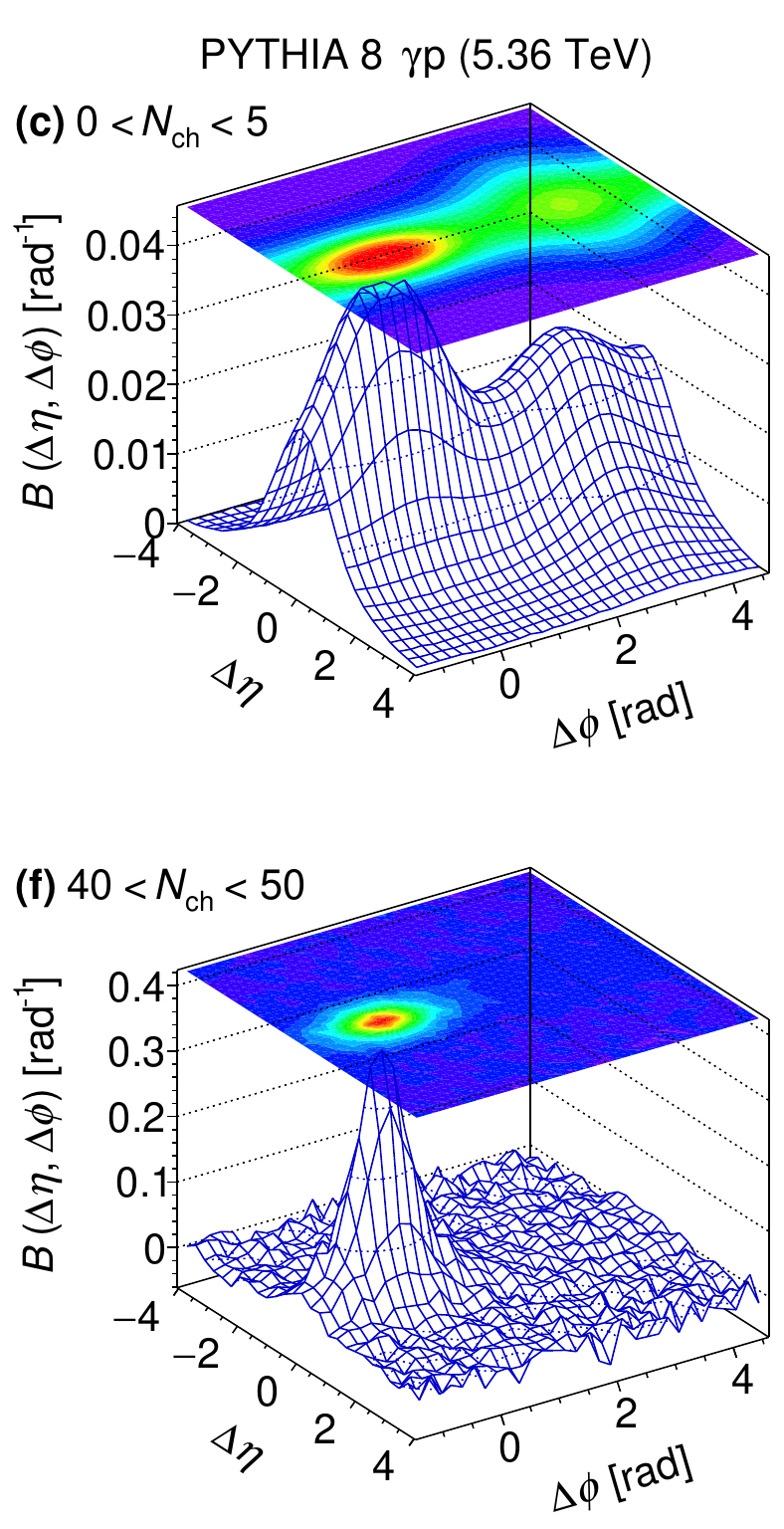}
 
 \caption{Two-dimensional number balance functions ($B$) for $0 < N_{\mathrm{ch}} < 5$ (upper panel) and $40 < N_{\mathrm{ch}} < 50$ (lower panel) of charged hadrons in \textsc{mb}-like tune pp (left), \textsc{mb}-like tune \gp (middle), and \gp (right) collisions at $\sqrt{s}=5.36$ TeV from \pythia. The results are obtained in the kinematic region $|\eta| < 2.4$ and $0.3 < p_{\mathrm{T}} < 3.0$ GeV.}
\label{fig:BF2D_comp}
\end{figure*}

Figure~\ref{fig:BF2D_comp} presents the two-dimensional balance function $B$ for charged hadrons in terms of $\Delta \eta$ and $\Delta \phi$ for two different multiplicity intervals in \gp and pp collisions at $\sqrt{s}$ = 5.36 TeV. The low and high multiplicities correspond to $0 < N_\mathrm{ch} < 5$ (upper panels) and $40 < N_\mathrm{ch} < 50$ (lower panels), respectively. \\

At very low multiplicity ($0 < N_{\mathrm{ch}} < 5$) in pp and \textsc{mb}-like tune \gp collisions, the balance function exhibits a strong away-side correlation around $\Delta\phi \approx \pi$, reflecting the dominance of a single hard parton scattering and the resulting back-to-back dijet topology. As the multiplicity increases, this away-side structure decreases in magnitude due to the growing importance of MPI, while the nearside peak at $\Delta\eta \approx 0$ becomes significantly stronger, rising from about $0.01$ at $\nch$ $\approx$ 5 to approximately $0.1$ for $40 < \nch < 50$. This behavior is consistent with a transition from long-range charge balancing across dijet events at low multiplicity to predominantly local charge conservation (LCC) at higher multiplicity. \\

However, the results from the \gp collisions (top, right) without \textsc{mb}-like tune of Fig.~\ref{fig:BF2D_comp} exhibit different behavior compared to the pp collisions at similar \nch.  At very low multiplicity, the $B$ shows a much weaker away-side peak, which rapidly disappears with increasing multiplicity. This suppression is attributed to the lack of MPI and hard QCD dijet contributions because the photon interacts predominantly through a single, highly asymmetric color string attached to a quark in the proton.\\

At the same time, the amplitude $B$ shows a stronger nearside peak in the \gp collisions than pp at the similar \nch. The amplitude changes approximately $0.04$ to 0.4 from low to high \nch. This enhanced nearside peak in \gp collisions originates from the fact that the charge production is governed almost exclusively by a single, short string with strong local charge ordering. In the absence of MPI, minijets, and multiple color strings, the charge-balancing partners remain tightly correlated in $\Delta\eta$ and $\Delta\phi$ phase space. \\

\begin{figure*} [htbp]  % CI Gamma p  collisions
    \centering
        \includegraphics[width=0.3\textwidth]{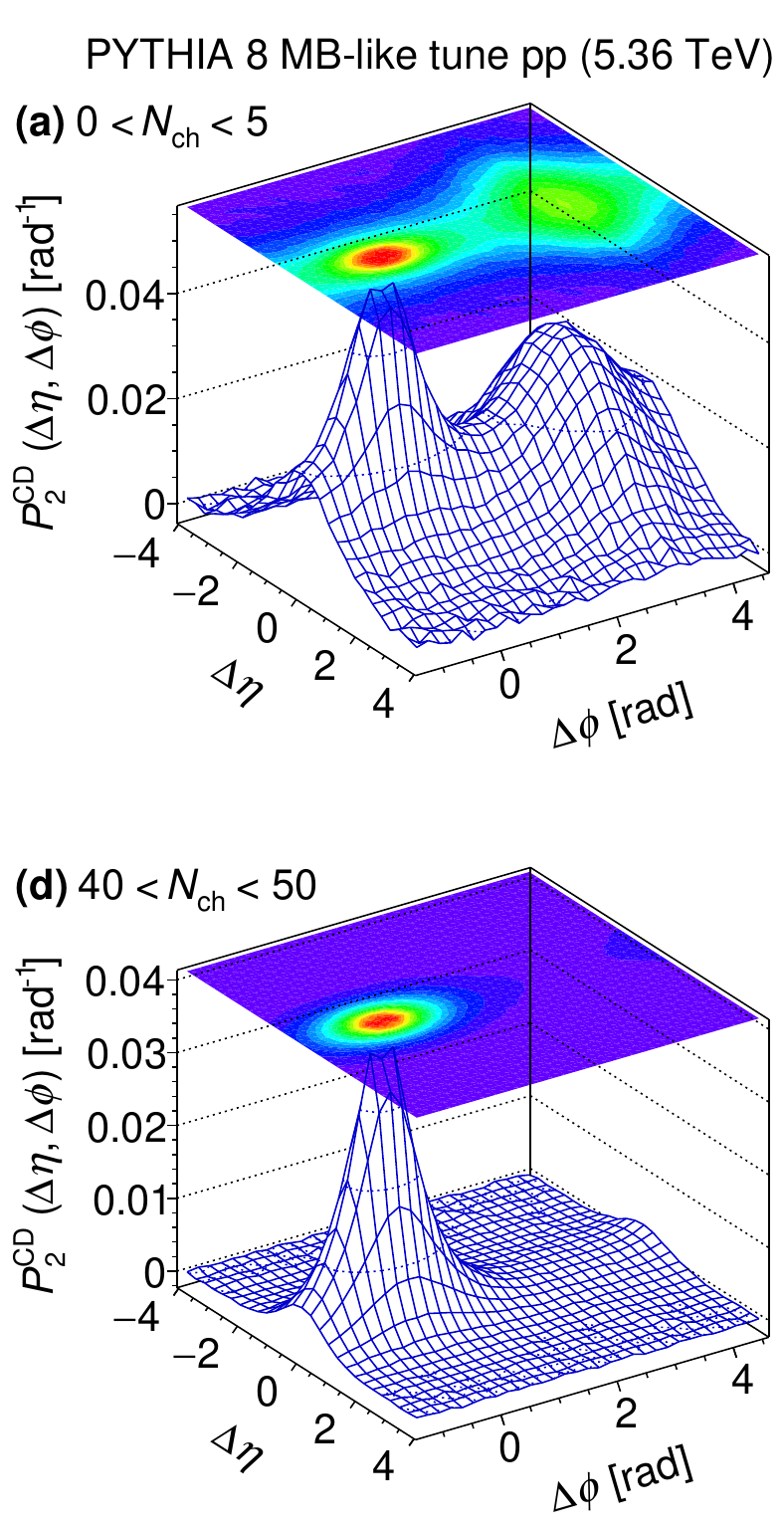}
        \includegraphics[width=0.3\textwidth]{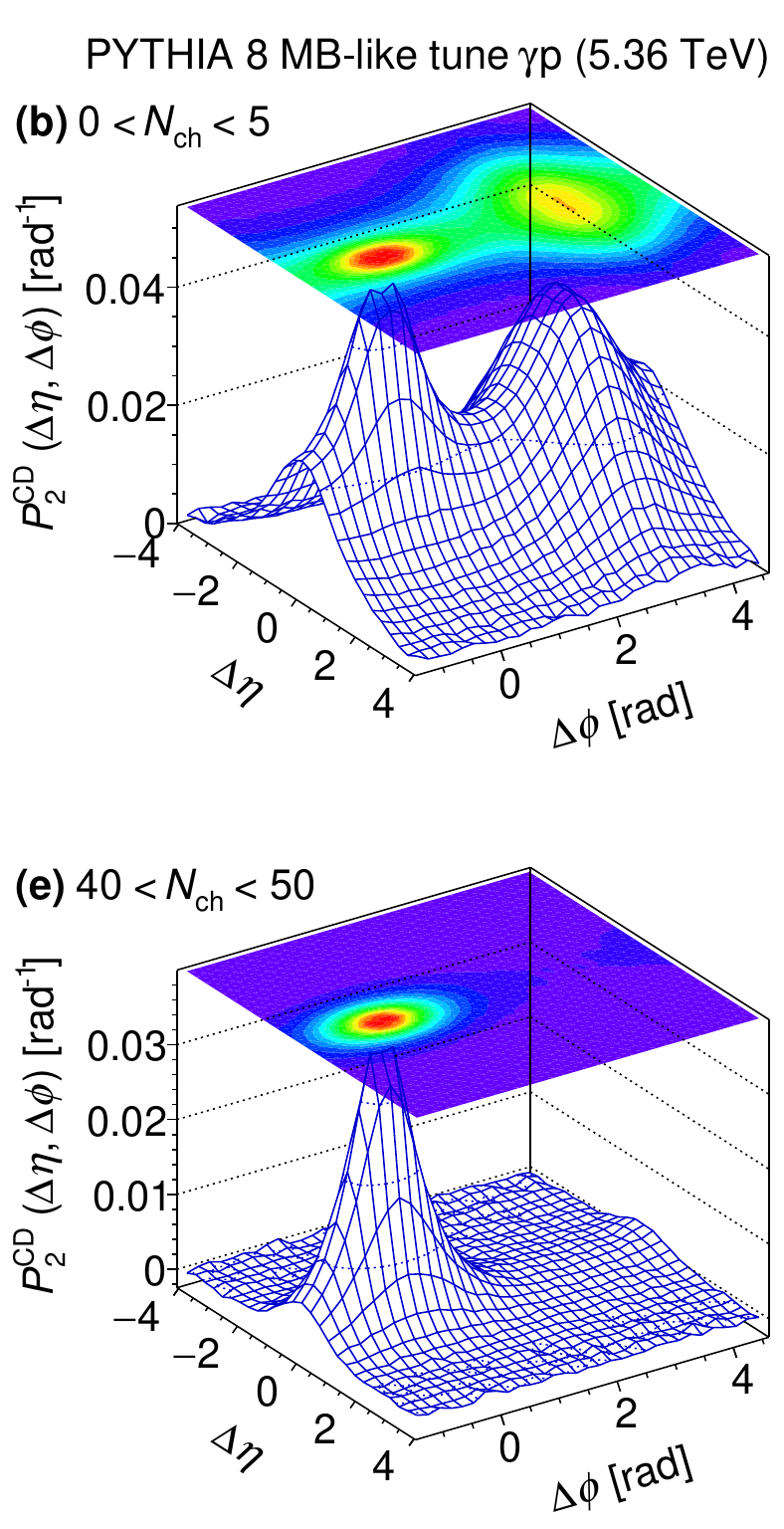}
        \includegraphics[width=0.3\textwidth]{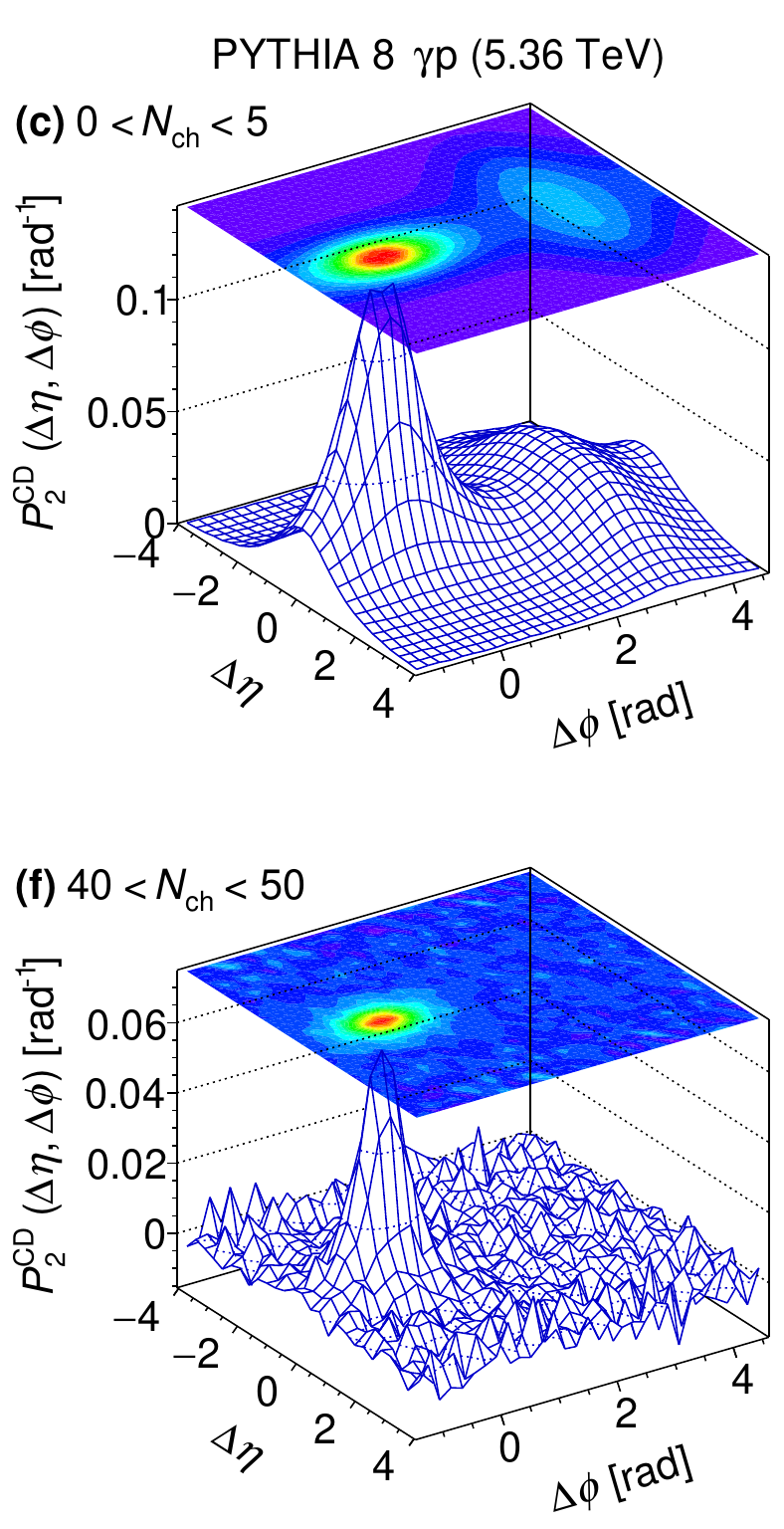}
\caption{Two-dimensional momentum balance functions ($P_{2}^\mathrm{CD}$) for $0 < N_{\mathrm{ch}} < 5$ (upper panel) and $40 < N_{\mathrm{ch}} < 50$ (lower panel) of charged hadrons in \textsc{mb}-like tune pp (left), \textsc{mb}-like tune \gp (middle), and \gp (right) collisions at $\sqrt{s}=5.36$ TeV from \pythia. The results are obtained in the kinematic region $|\eta| < 2.4$ and $0.3 < p_{\mathrm{T}} < 3.0$ GeV.}
\label{fig:momB_comp}
\end{figure*}  
Similarly, the two-dimensional charge-dependent momentum balance correlation $P_{2}^\mathrm{CD}(\Delta\eta,\Delta\phi)$ is shown in Fig. \ref{fig:momB_comp}. A distinct feature is seen when comparing pp and \gp collisions. At very low \nch, the nearside peak is significantly stronger in \gp interactions than in pp events. This enhancement originates from the single-string topology in \gp collisions, where opposite-charge particles are produced in a more localized manner along the string, resulting in tighter momentum correlations.\\

Interestingly, the away-side correlation $\Delta\phi \approx \pi$ is observed to have similar strength in both systems within the same multiplicity interval. As the event multiplicity increases toward $N_{\mathrm{ch}}\approx 50$, the nearside peak, the strengths converge to similar levels between pp and \gp collisions. The results from the \gp collisions with the \textsc{mb}-like tune exhibit qualitatively similar behavior to that observed in the pp collisions. It is found that the $B$ and $P_{2}^{\mathrm{CD}}$ observables show similar multiplicity dependence, indicating that they depend more strongly on the underlying processes than the nature of the colliding system.\\

Overall, a narrower nearside structure in $P_{2}^{\mathrm{CD}}$ relative to $B$ indicates that balancing charges originating from harder partonic processes, such as jet fragmentation, exhibit stronger localized momentum correlations. This difference highlights the sensitivity of these observables to both the topology and the dynamical properties of particle production.\\

In addition to the CD correlations, a brief discussion about the CI correlations is provided in Sec~\ref{append} (Appendix).

\subsection{One-dimensional projections}
\begin{figure*}  % Wide figure spanning both columns
    \centering
     \subfigure[]{
        \includegraphics[width=0.3\textwidth]{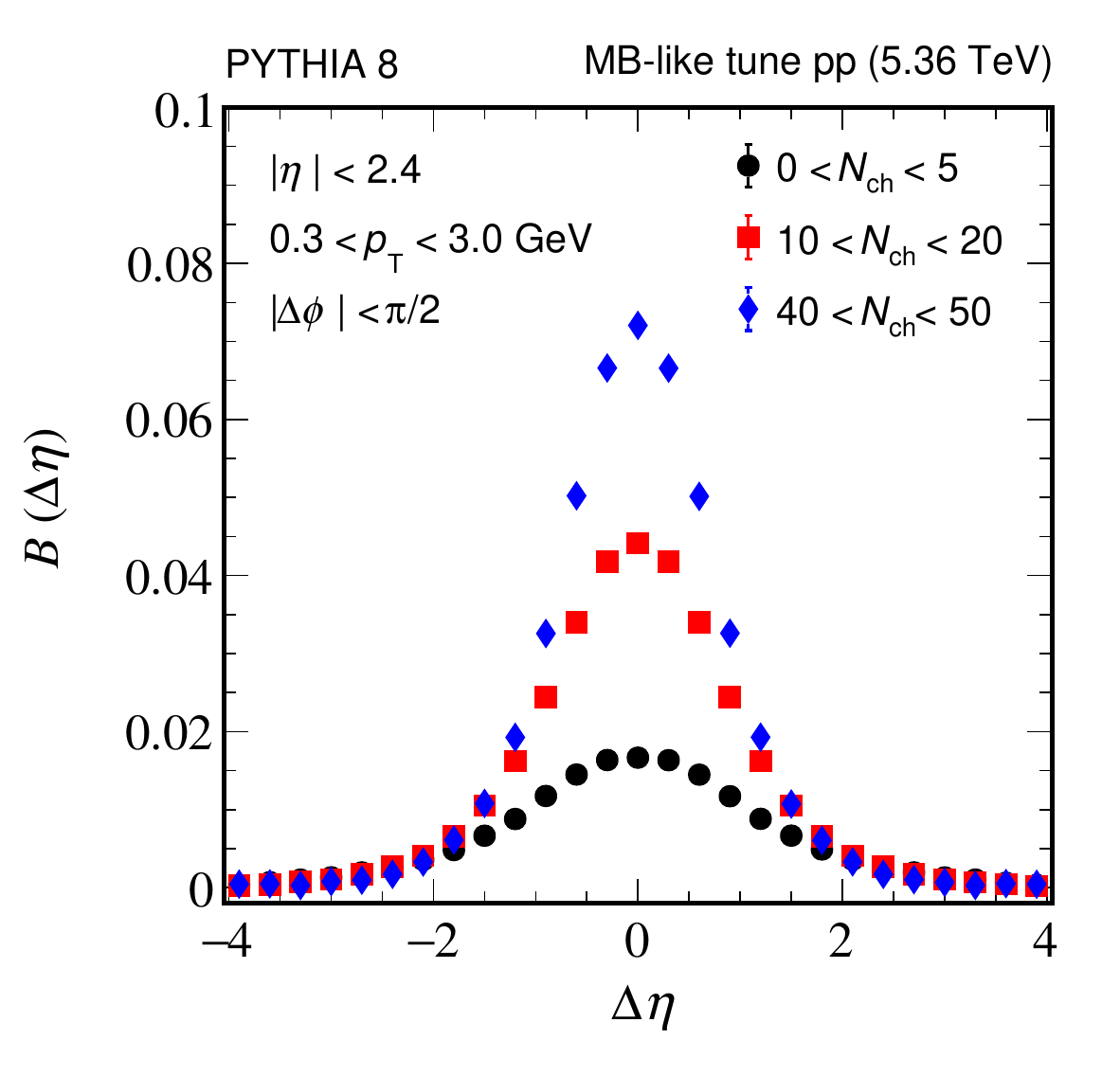}
    }
    \subfigure[]{
        \includegraphics[width=0.3\textwidth]{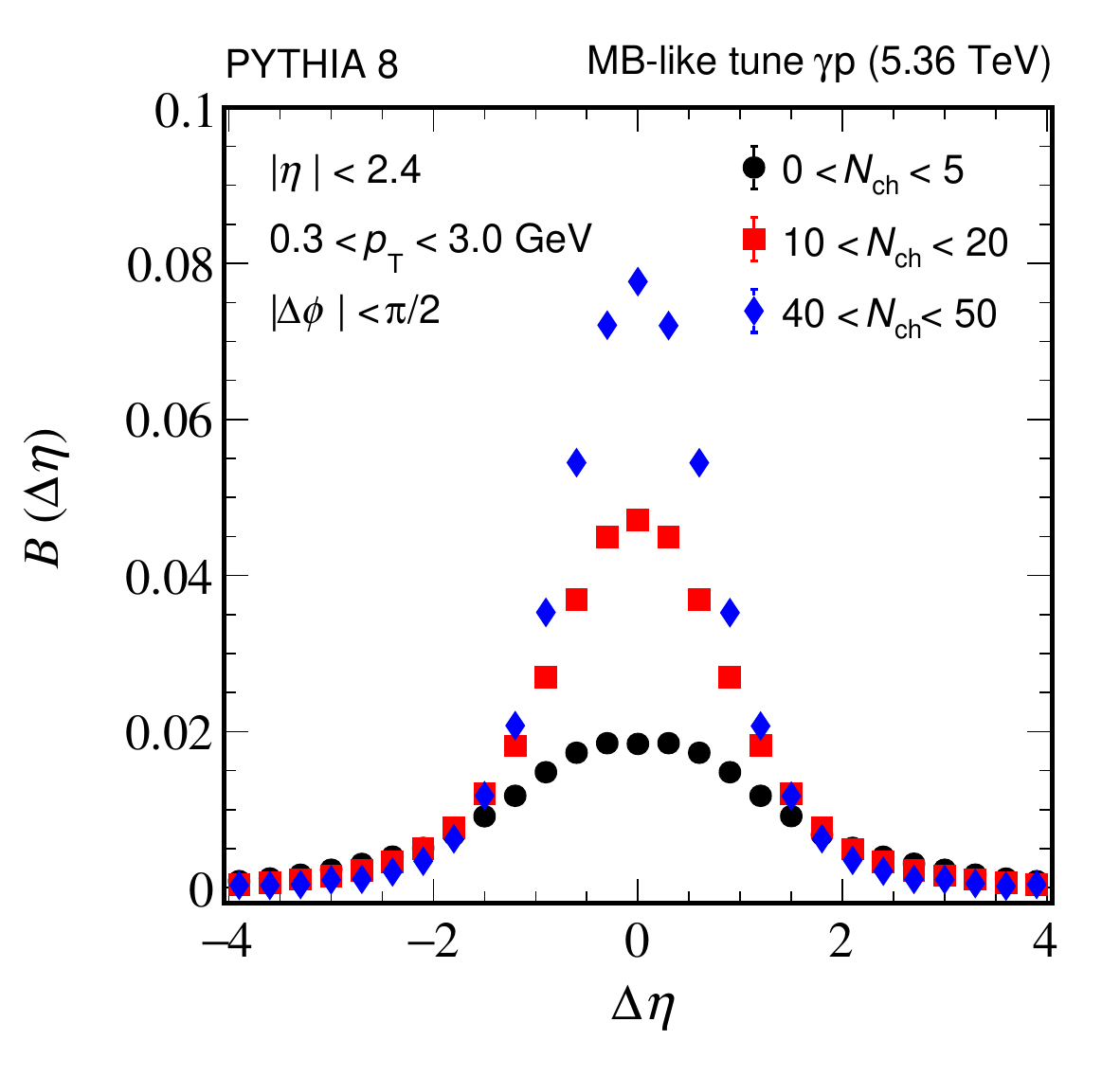}
    }
    \subfigure[]{
        \includegraphics[width=0.3\textwidth]{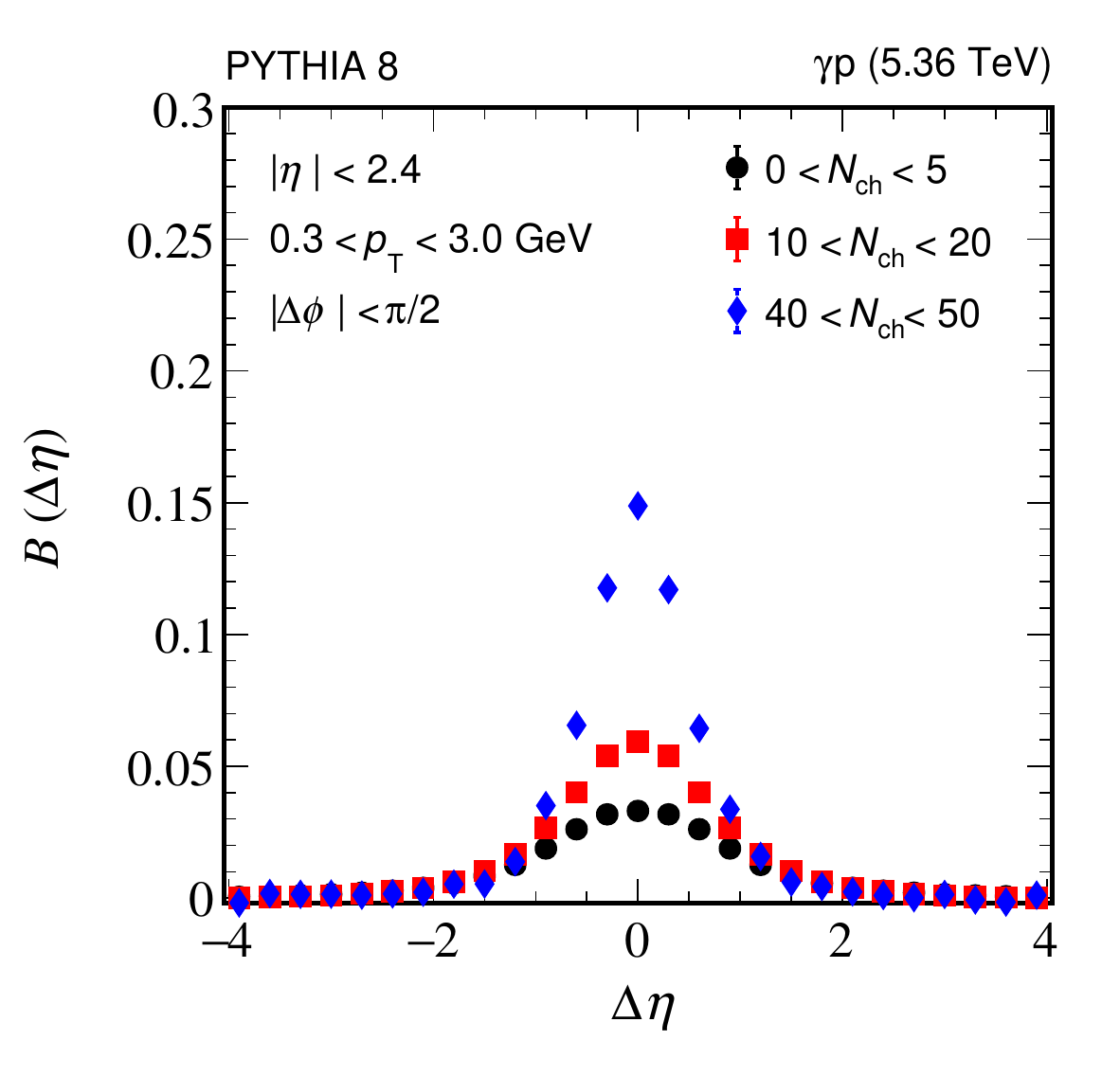}
    }
    \subfigure[]{
        \includegraphics[width=0.3\textwidth]{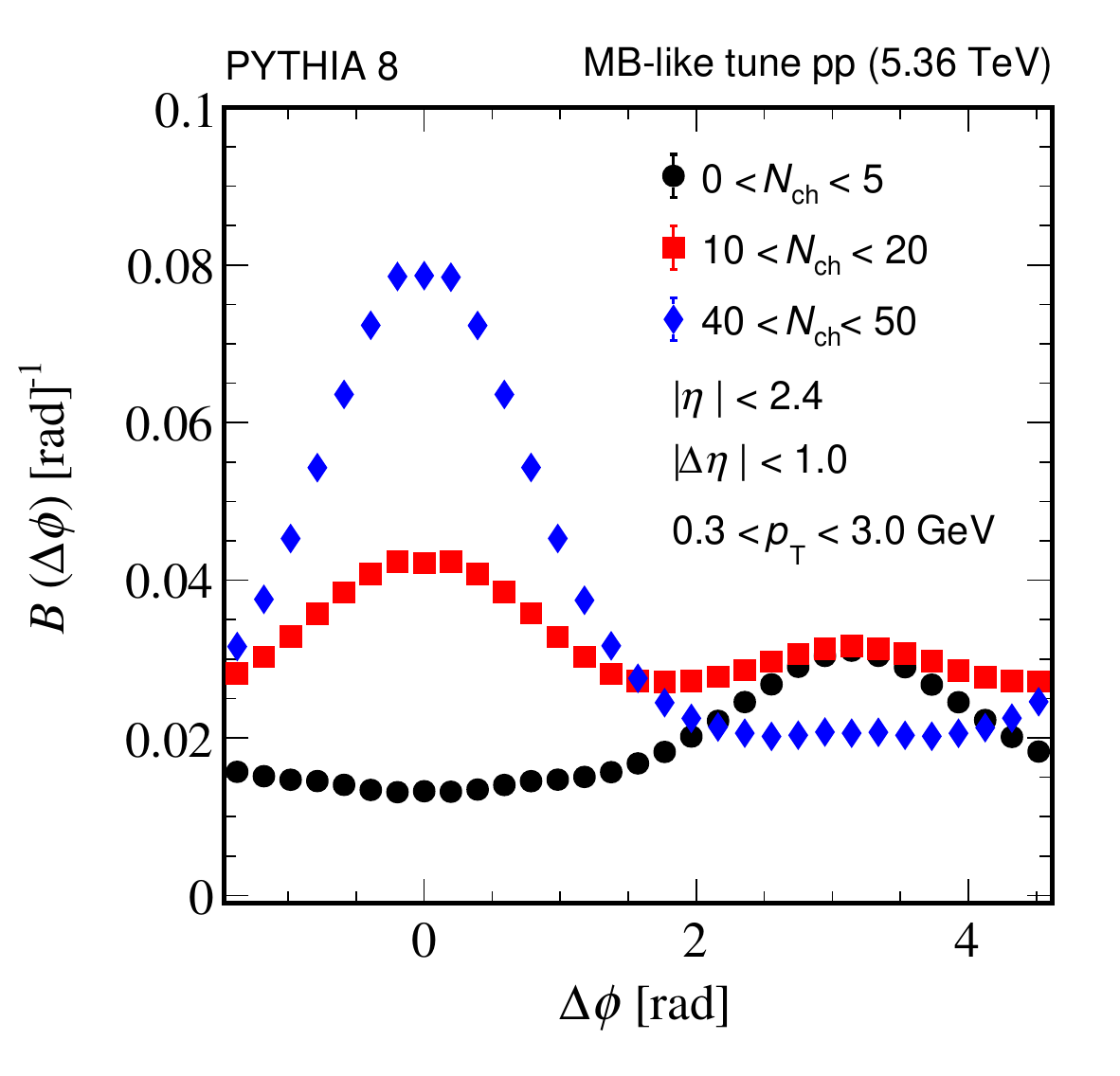}
    }
    \subfigure[]{
        \includegraphics[width=0.3\textwidth]{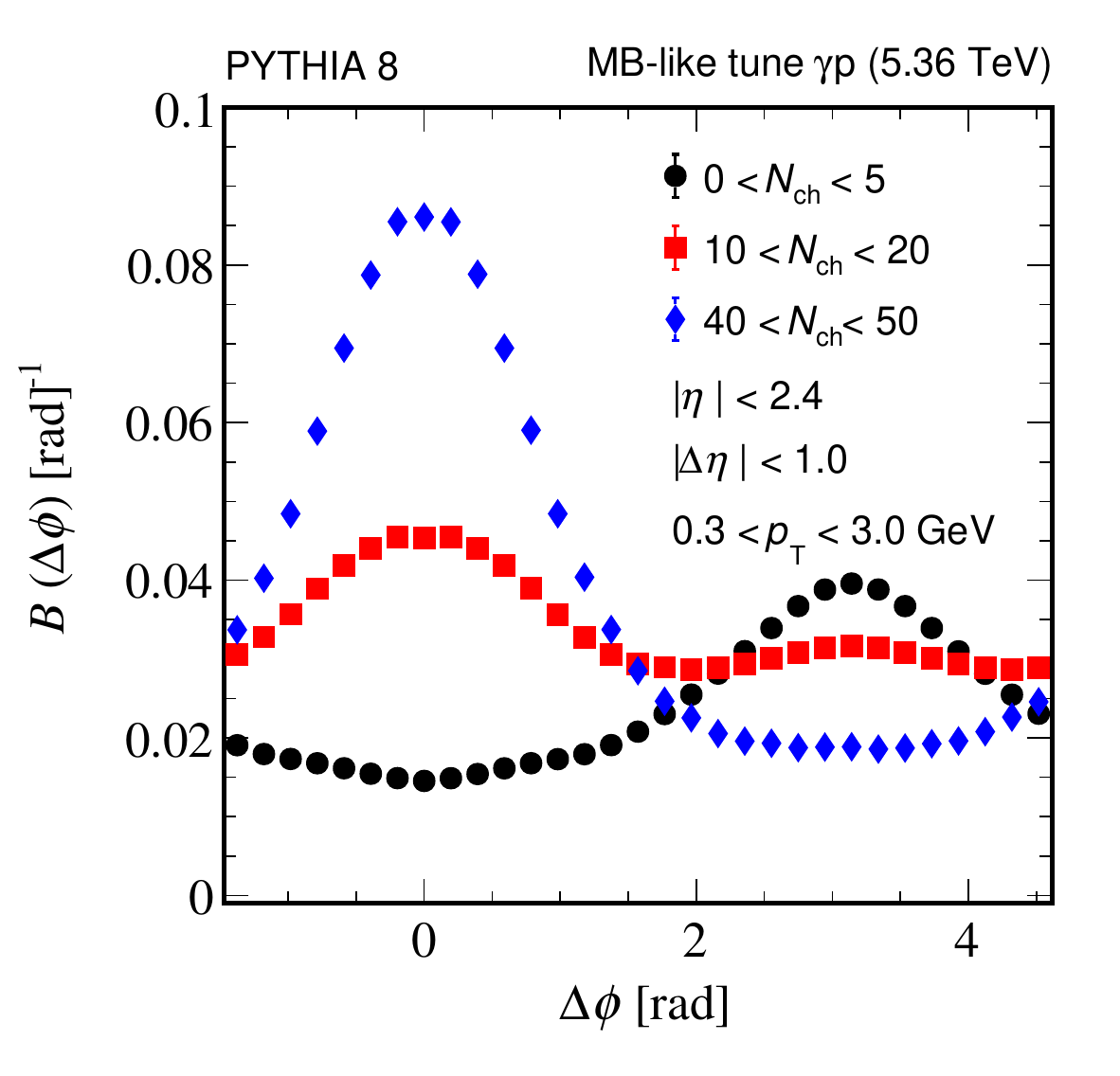}
    }
    \subfigure[]{
        \includegraphics[width=0.3\textwidth]{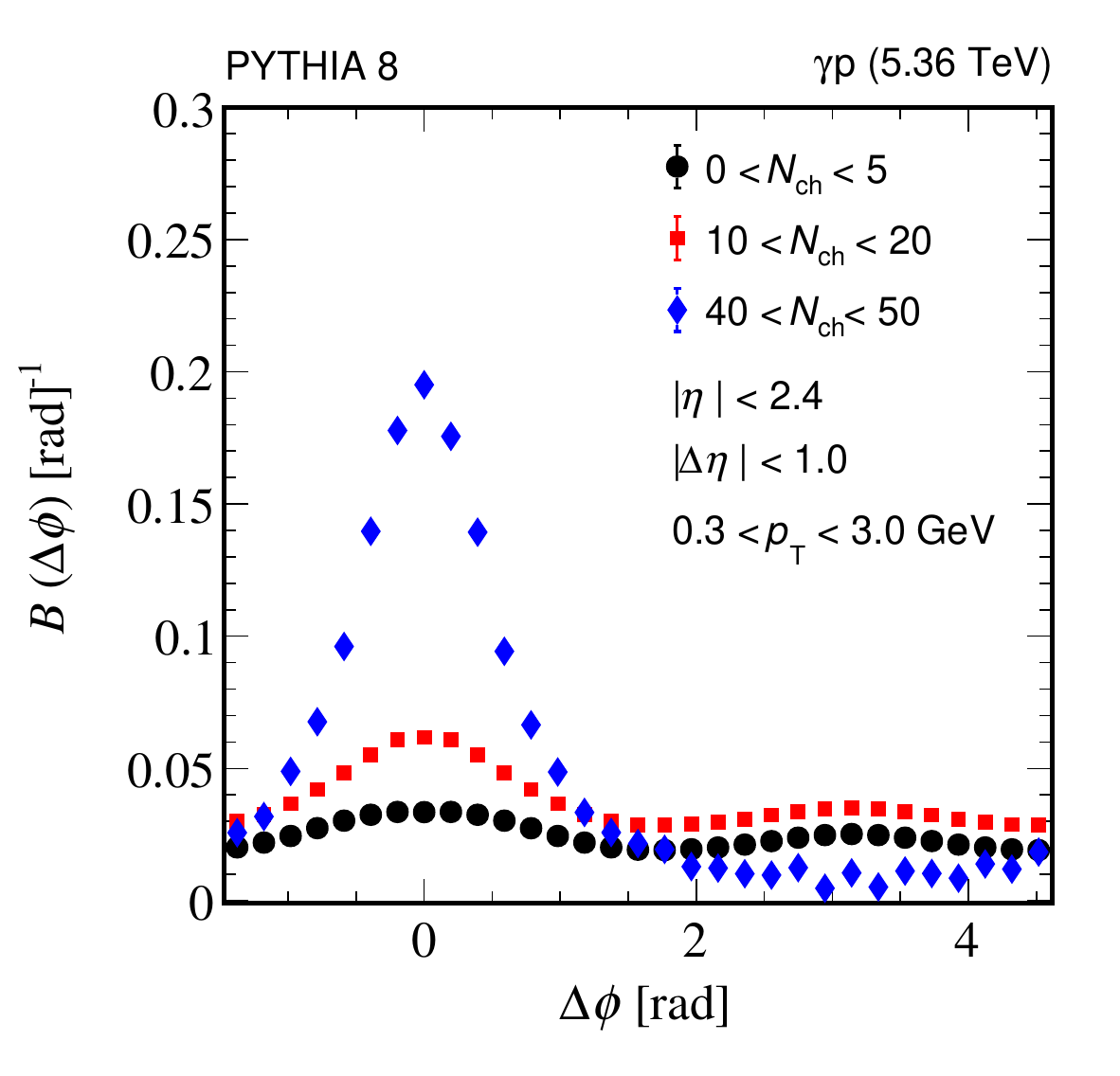}
    }
    \caption{One-dimensional nearside projection of the number balance functions ($B$) as a function of $\Delta\eta$ (upper panel) and $\Delta\phi$ (lower panel) in \textsc{mb}-like tune pp (left), \textsc{mb}-like tune \gp (middle), and \gp (right) collisions at $\sqrt{s}=5.36$ TeV from \pythia. The results are obtained in the kinematic region $|\eta| < 2.4$ and $0.3 < p_{\mathrm{T}} < 3.0$ GeV for charged hadrons.}
    \label{fig:model_plots_1d_B}
\end{figure*}

A clear multiplicity dependence is observed in the 1D distributions of $B$ as a function of $\Delta\eta$ and $\Delta\phi$ for three \nch intervals in pp and $\gamma$p collisions, as shown in Fig.~\ref{fig:model_plots_1d_B}. The nearside peak of the 1D $B$ distribution becomes narrower with increasing \nch. This narrowing is more pronounced in \gp compared to pp collisions. Interestingly, the near sidepeak of $B(\Delta\phi)$ is nearly flat in pp, but a smaller peaklike structure is seen for the \gp at the low \nch region. However, at higher \nch,  a sharper peak is found in \gp than pp collisions. \\

%%%%%% P2CD plot %%%%

\begin{figure*}  % Wide figure spanning both columns
    \centering
     \subfigure[]{
        \includegraphics[width=0.3\textwidth]{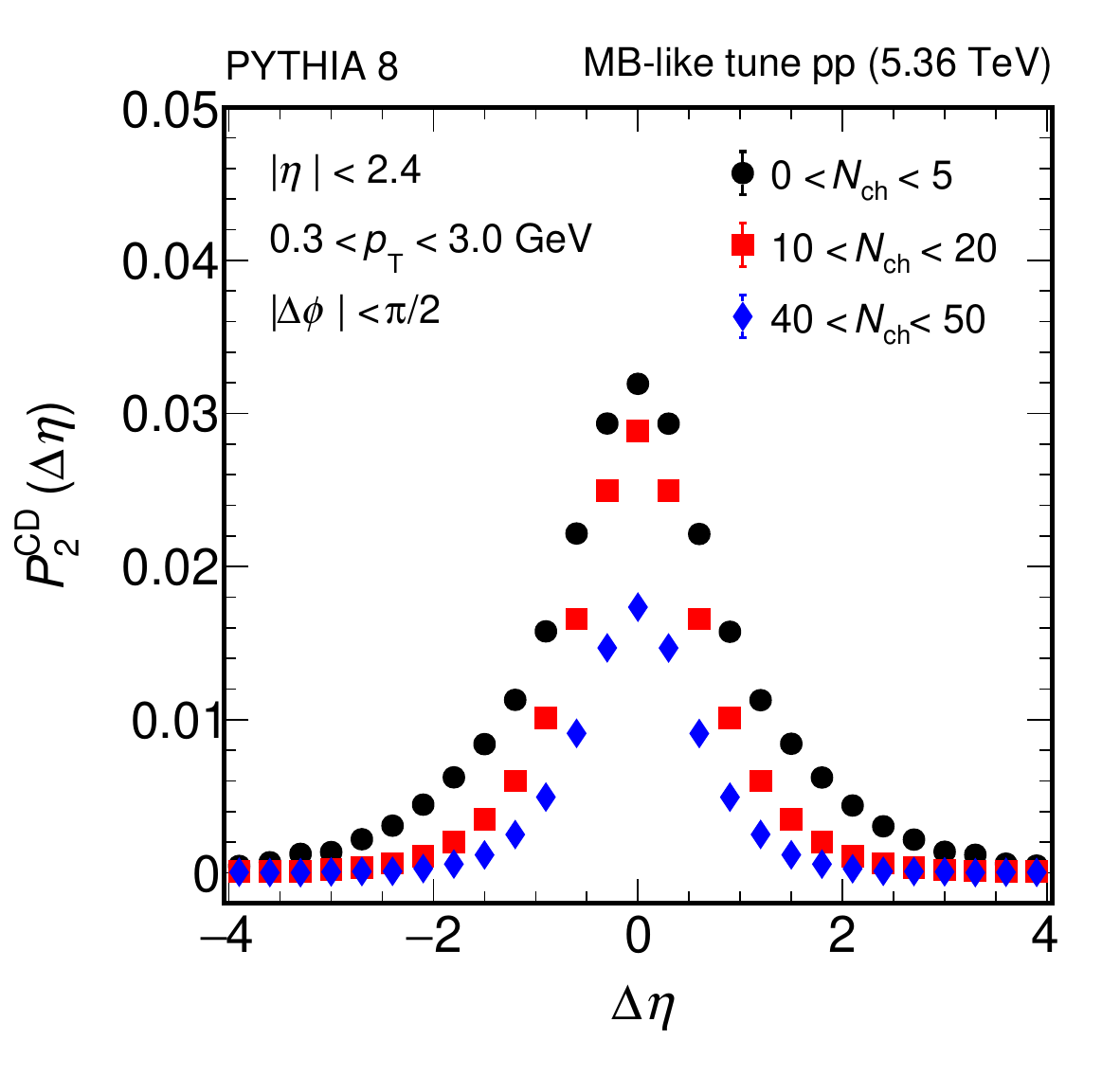}
    }
    \subfigure[]{
        \includegraphics[width=0.3\textwidth]{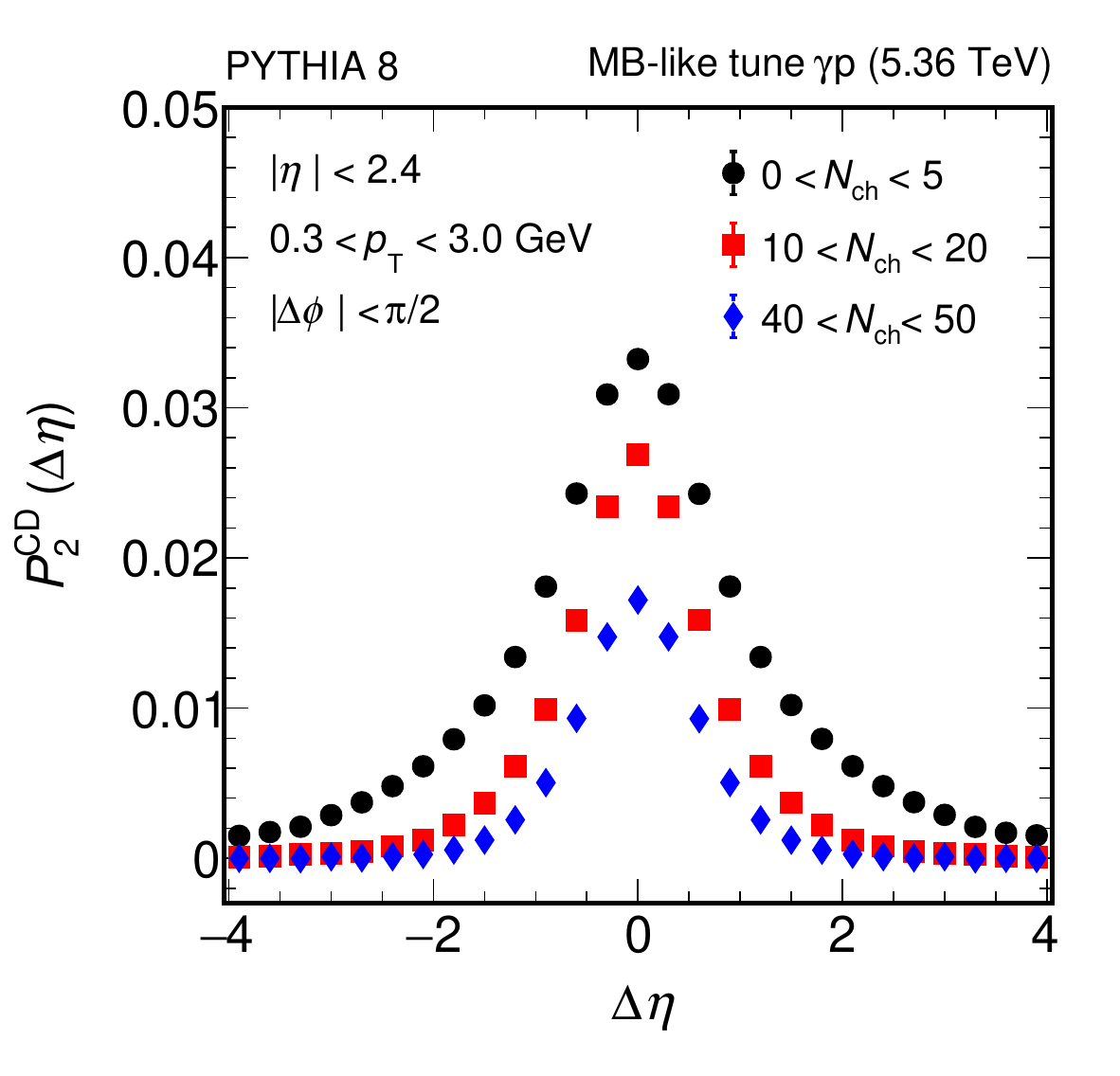}
    }
    \subfigure[]{
        \includegraphics[width=0.3\textwidth]{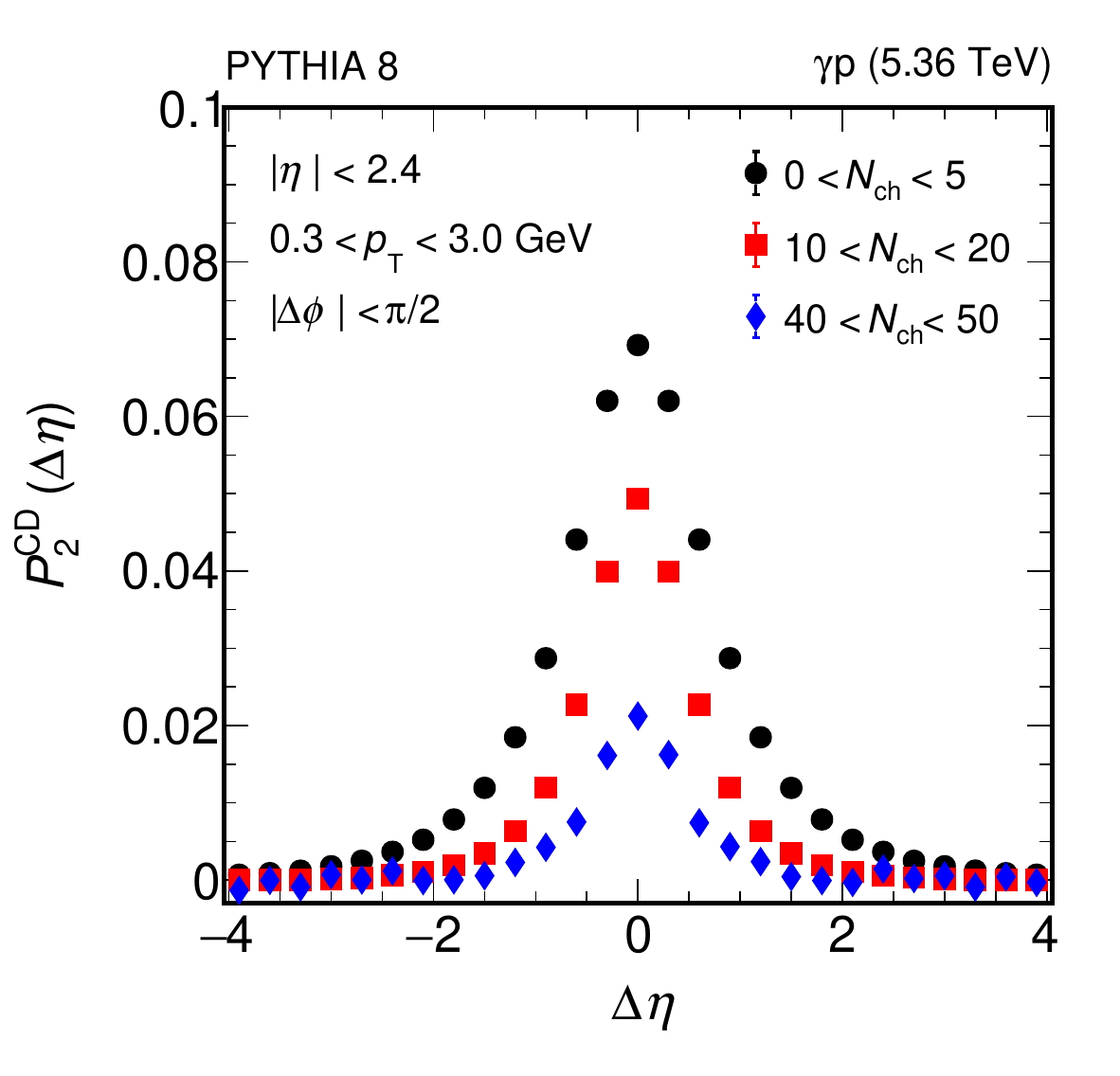}
    }
    \subfigure[]{
        \includegraphics[width=0.3\textwidth]{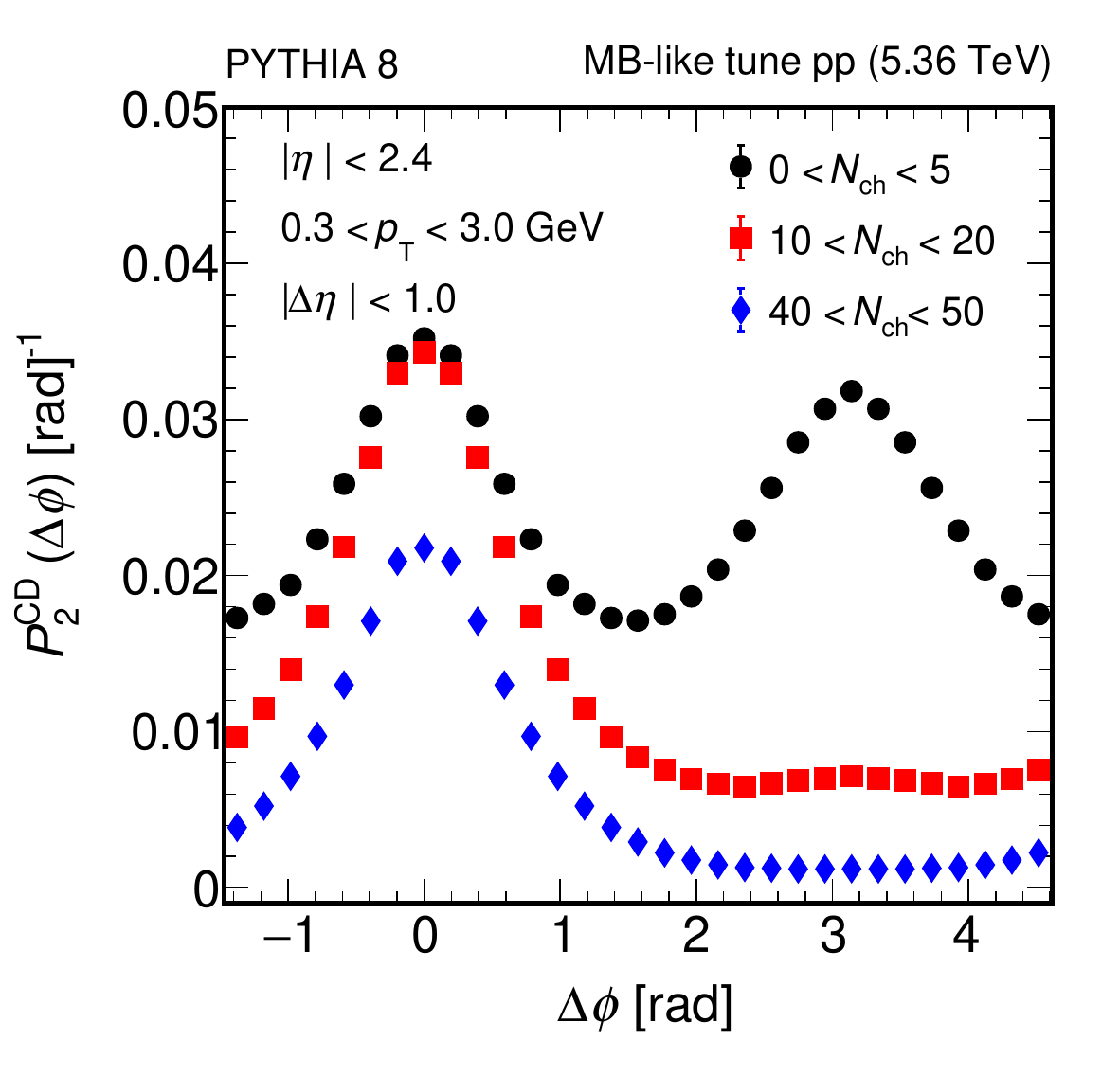}
    }
    \subfigure[]{
        \includegraphics[width=0.3\textwidth]{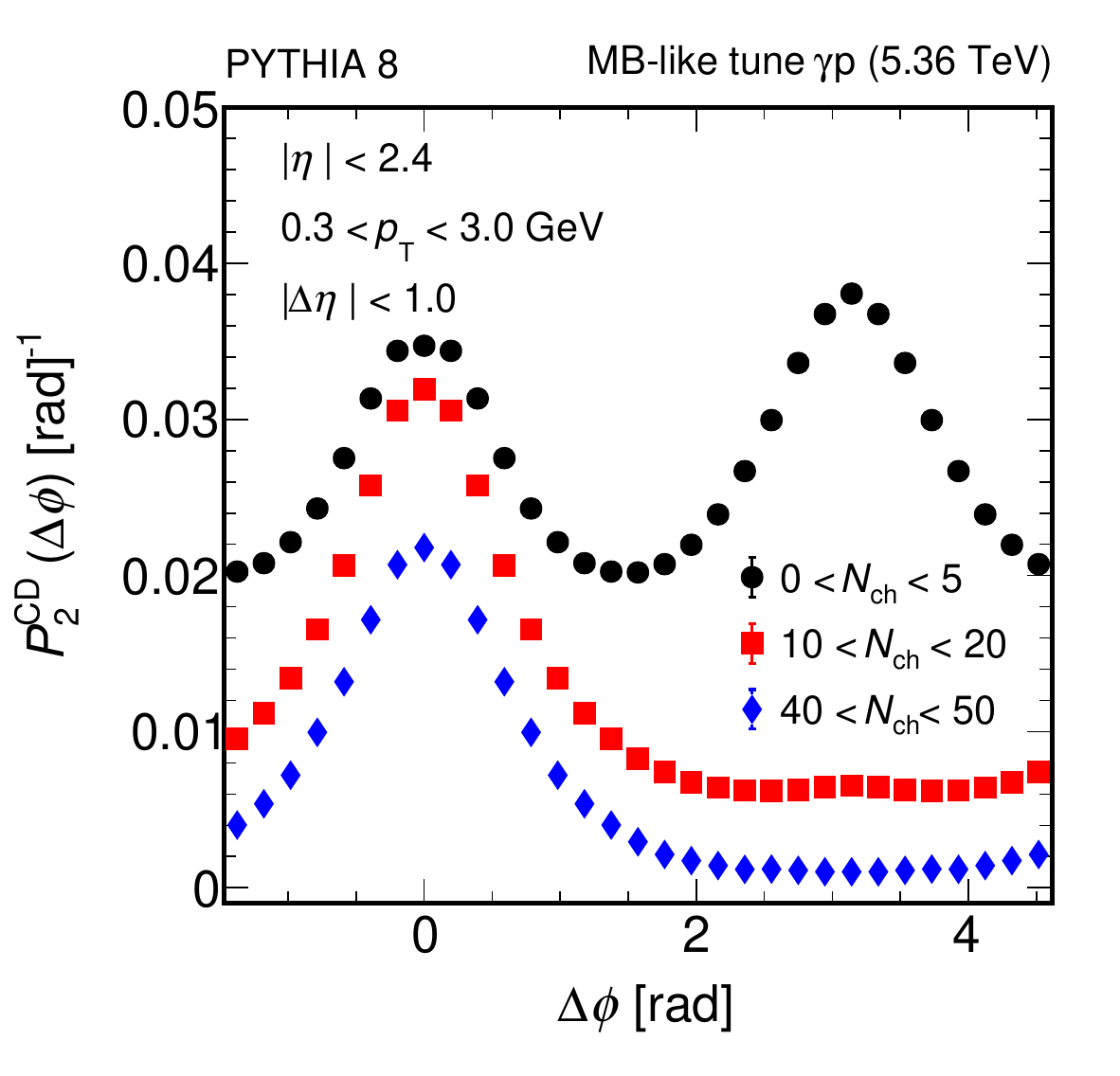}
    }
    \subfigure[]{
        \includegraphics[width=0.3\textwidth]{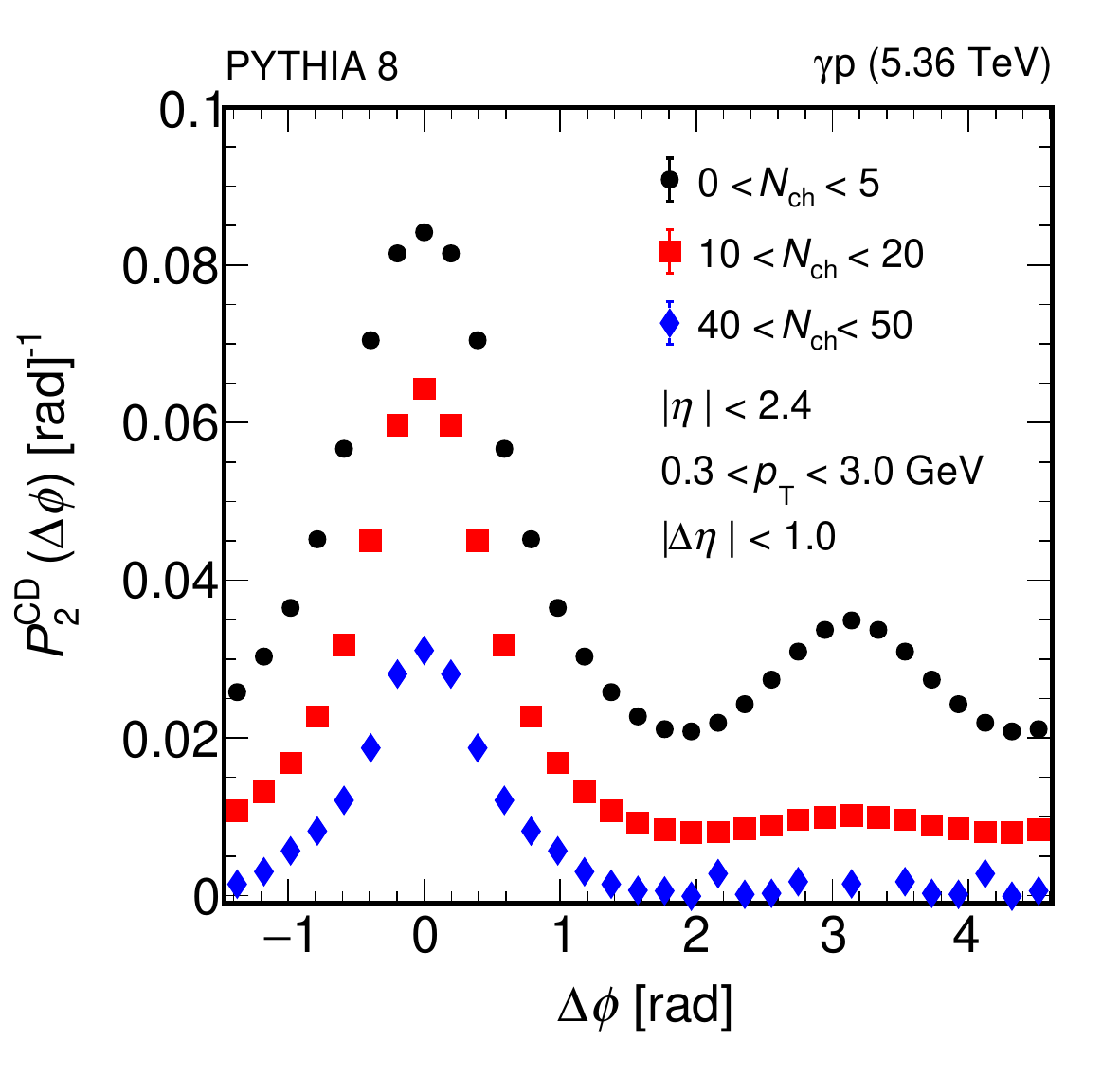}
    }
    \caption{One-dimensional nearside projection of the momentum balance functions ($P_{2}^\mathrm{CD}$) as a function of $\Delta\eta$ (upper panel) and $\Delta\phi$ (lower panel) in \textsc{mb}-like tune pp (left), \textsc{mb}-like tune \gp (middle), and \gp (right) collisions at $\sqrt{s}=5.36$ TeV from \pythia. The results are obtained in the kinematic region $|\eta| < 2.4$ and $0.3 < p_{\mathrm{T}} < 3.0$ GeV for charged hadrons.}
    \label{fig:model_plots_1d_p2cd}
\end{figure*}

Figure~\ref{fig:model_plots_1d_p2cd} presents the 1D projections of $P_{2}^\mathrm{CD}$ along $\Delta\eta$ and $\Delta\phi$ directions. $P_{2}^\mathrm{CD}$ shows a clear multiplicity evolution, and their amplitudes gradually decrease with increasing \nch.  This can be attributed to the combined contribution from \mpt weighting and jet-dominated processes, which is more prominent in low-multiplicity pp events.
The observed trend in the $P_{2}^\mathrm{CD}$ is opposite to the $B$ and the difference arises between them, because $P_{2}^\mathrm{CD}$ is not only sensitive to the number of the correlated particle pairs but also to their \pt correlations.

\subsection{Width comparison}

\begin{figure*}  % Wide figure spanning both columns
    \centering
 \includegraphics[width=1.0\textwidth]{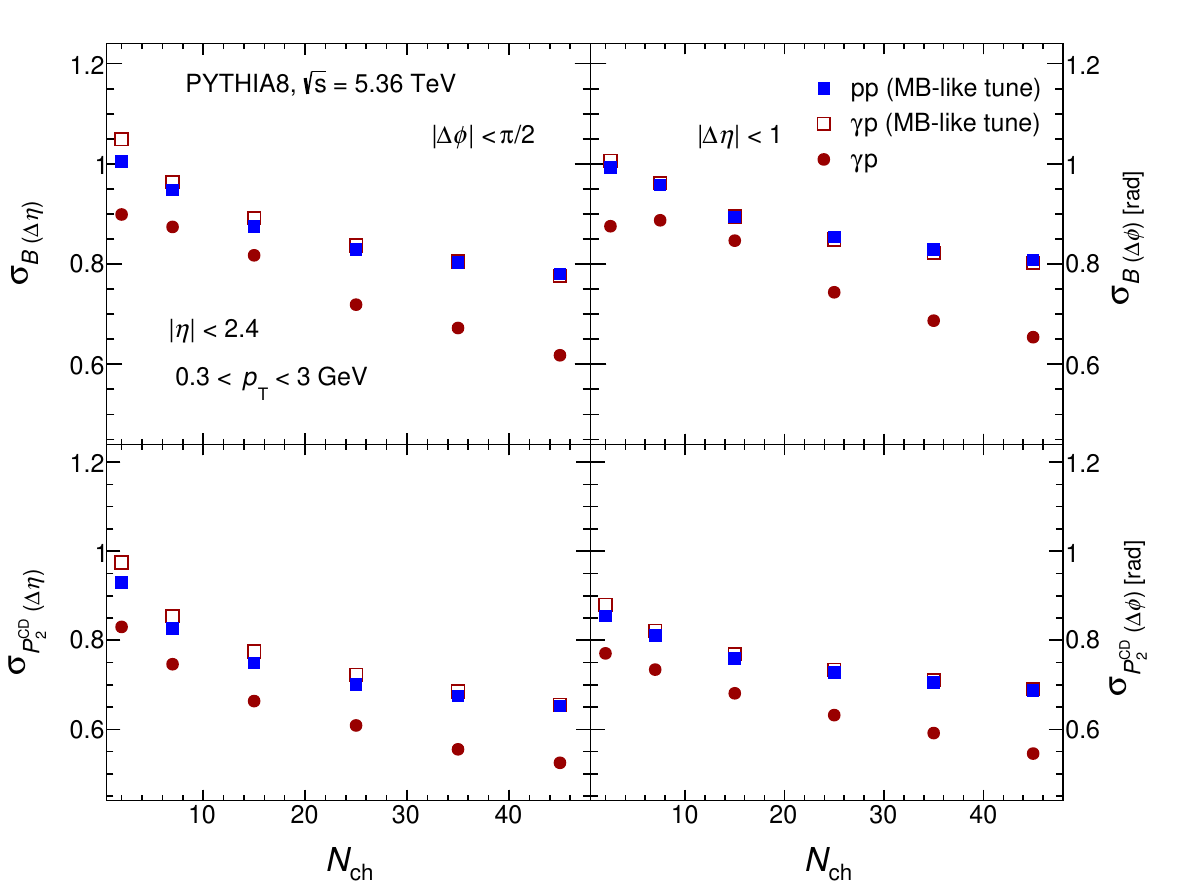}
    \caption{The RMS width of the number ($\sigma_{B}$) and momentum ($\sigma_{P_{2}^\mathrm{CD}}$) of balance functions in terms of $\Delta\eta$ (left column) and $\Delta\phi$ (right column) with \nch. Both the trigger and associated particles are considered in $0.3 < p_\mathrm{T} < 3.0$ GeV and $|\eta| < 2.4$ for the charged hadrons.}
    \label{fig:width_plots}
  \end{figure*}

The strength of the correlation between produced hadron pairs is reflected in the width of the balance functions and is quantified by the observables $\sigma(\Delta\eta)$ and $\sigma(\Delta\phi)$, as defined in Eq.~\ref{eq:BFwidth}. Figure~\ref{fig:width_plots} shows a clear multiplicity-dependent trend: the widths of $B$ and $P_{2}^\mathrm{CD}$ decrease with increasing \nch for both pp and \gp collisions. The trends observed in this study are qualitatively consistent with the available experimental measurements for pp collisions by the ALICE Collaboration~\cite{alicethree}, despite differences in the kinematic selections, multiplicity definitions, and collision energies. However, both results show similar qualitative behavior on the multiplicity-dependent study of the balance function for charged hadrons. Future measurements in small systems and photon-induced processes at the LHC energies will provide valuable experimental constraints and help further assess the underlying mechanisms responsible for the observed behavior. \\

The extracted widths of these correlations are smaller in \gp collisions than in pp collisions across all multiplicity classes, with the difference becoming more pronounced at higher \nch,

\[
\sigma^{\gamma p}_{P_{2}^\mathrm{CD}(\Delta\eta,\Delta\phi) ~\text{or} ~B(\Delta\eta,\Delta\phi)} < \sigma^{pp}_{P_{2}^\mathrm{CD}(\Delta\eta,\Delta\phi) ~\text{or} ~B(\Delta\eta,\Delta\phi)}.
\]
 
A comparison of the widths of the $B$ and $P_{2}^\mathrm{CD}$ reveals a clear and systematic dependence on the charged particle multiplicity. The ordering of the widths is preserved across all multiplicity classes,
\[
\sigma_{P_{2}^\mathrm{CD}(\Delta\eta,\Delta\phi)} < \sigma_{B(\Delta\eta,\Delta\phi)},
\]
suggesting that the underlying mechanisms of charge conservation for the balance function and localized jetlike momentum correlations for $P_2$ remain robust across collision systems.
In addition, no significant difference in the extracted widths is observed between \gp and pp collisions when the same \pythia configuration is used, despite the different colliding systems. However, a small difference appears at low \nch. 
These findings highlight a clear picture of multiplicity-driven narrowing in both pp and \gp collisions, while this also points to open questions regarding the relative strength of MPI, color reconnection, and hard processes in photon-induced interactions and how these mechanisms scale with event activity in small systems.\\

To further investigate the sensitivity of balance-function observables to different model configurations, additional studies were performed by varying the CR, MPI, and soft-QCD processes relative to the default configuration used in this study. \\

\begin{figure*}  % Wide figure spanning both columns
    \centering
         \includegraphics[width=1.0\textwidth]{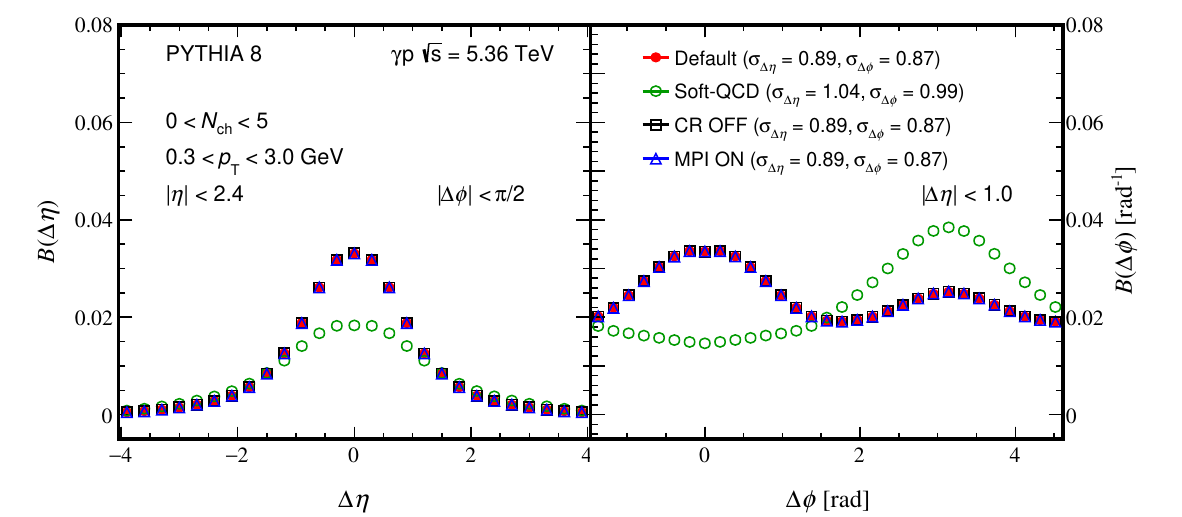}
        \includegraphics[width=1.0\textwidth]{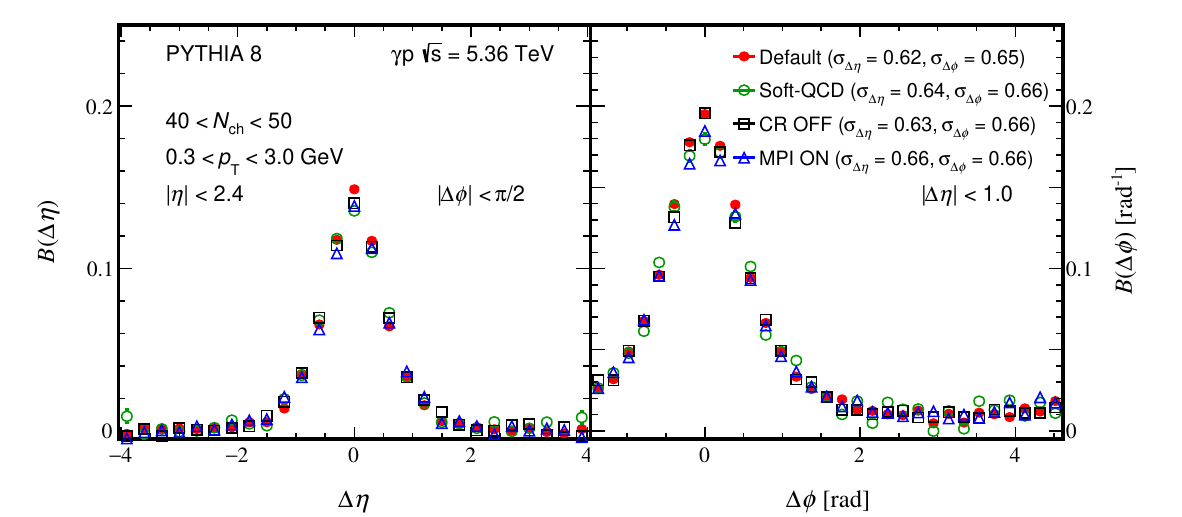}

    \caption{Balance functions comparison using different \pythia configurations. The upper (lower) panels correspond to the multiplicity range $0 < N_\mathrm{ch} < 5$ ($40 < N_\mathrm{ch} < 50$). The left (right) column presents the $\Delta\eta$ ($\Delta\phi$) balance functions in $0.3 < p_\mathrm{T} < 3.0$ GeV and $|\eta| < 2.4$.}
    \label{pythia_diffTune}
\end{figure*}

Figure~\ref{pythia_diffTune} presents a comparison of the balance functions $B(\Delta\eta)$ and $B(\Delta\phi)$ for low- and high-multiplicity events in \gp collisions. For the low-multiplicity interval ($0 < N_\mathrm{ch} < 5$), the soft-QCD-dominated sample exhibits significantly broader balance-function distributions compared to the default one. On the other hand, the CR-OFF and MPI-ON samples remain close to the default prediction. This indicates that soft particle-production processes have a stronger impact on the correlation widths than the individual CR or MPI enabled. At higher multiplicities ($40 < N_\mathrm{ch} < 50$), the differences among the various model configurations have smaller effects. A narrower width of the balance functions is found for all the configurations at higher \nch as compared to lower \nch. The similar behavior observed across different \pythia configurations at high multiplicity indicates that the narrowing of the balance function is associated with event multiplicity rather than with specific generator tunings used to model particle production mechanisms. However, as this study explores with a limited set of \textsc{pythia} model configurations, these results do not imply that multiplicity is the sole origin of the narrowing behavior but rather are suggestive of its role as a primary contributor.

\section{Summary}
\label{summ}
We present a study of two-particle number ($B$) and transverse-momentum ($P_{2}^\mathrm{CD}$) balance functions of inclusive charged particles as a function of event multiplicity in \gp collisions at LHC energies using \pythia simulations, and compare the results with pp collisions in the similar multiplicity range. Different \pythia configurations are considered, including setups with and without inelastic minimum-bias processes, to quantify the role of multiple parton interactions (MPIs), string-fragmentation based hadronization and underlying event activities on particle production.\\

The widths of both correlation observables decrease with increasing multiplicity in both pp and \gp collisions, demonstrating a clear multiplicity-dependent narrowing trend. Although photon-induced \gp events populate a lower multiplicity range than hadronic pp interactions, the same trend is present. The comparison at given multiplicity isolates the contribution of hard scattering. The low-multiplicity pp events exhibit a pronounced away-side peak driven by back-to-back dijet production, while such contributions are strongly suppressed in \gp collisions.
A dedicated study with identical minimum-bias configurations for pp and \gp collisions shows consistent results, suggesting that the observed multiplicity dependence is closely related to underlying event activity rather than the colliding system and the choice of specific physics processes.\\

This behavior in \gp interactions highlights the importance of future experimental measurements to validate these multiplicity-dependent effects. Future experimental measurements of photon-induced (i.e.,
 \gp, $\gamma$Pb, $\gamma$O) collisions  will therefore provide a clean environment to separate bulk-dominated nearside correlations at low transverse momentum from jet-driven away-side correlations characteristic of low-multiplicity pp events, delivering a robust baseline for disentangling soft and hard components of particle production across different collision systems.

~~~~~~~~~~~~~~~~~~~~~~~~~~~
~~~~~~~~~~~~~~~~~~~~~~~~~~~~~~~~~~~~
\begin{acknowledgements}
S. C. Behera and D. Mallick acknowledge the support under the Istituto Nazionale di Fisica Nucleare (INFN) Postdoctoral Fellowship Grant No. B. C. n. 25865 and No. B.C. n. 27077.
\end{acknowledgements}
\vspace{3cm}

%----------------------------------------------------------------
\bibliographystyle{utphys}   % Remember we use title in the biblio
\bibliography{BFppPythiaNotes}
%------------------------------------------------------------------------------------------------------------------------------------------
\section{Appendix: Charge-independent correlations}
\label{append}

\begin{figure*}  % CI correlations
    \centering
        \includegraphics[width=0.3\textwidth]{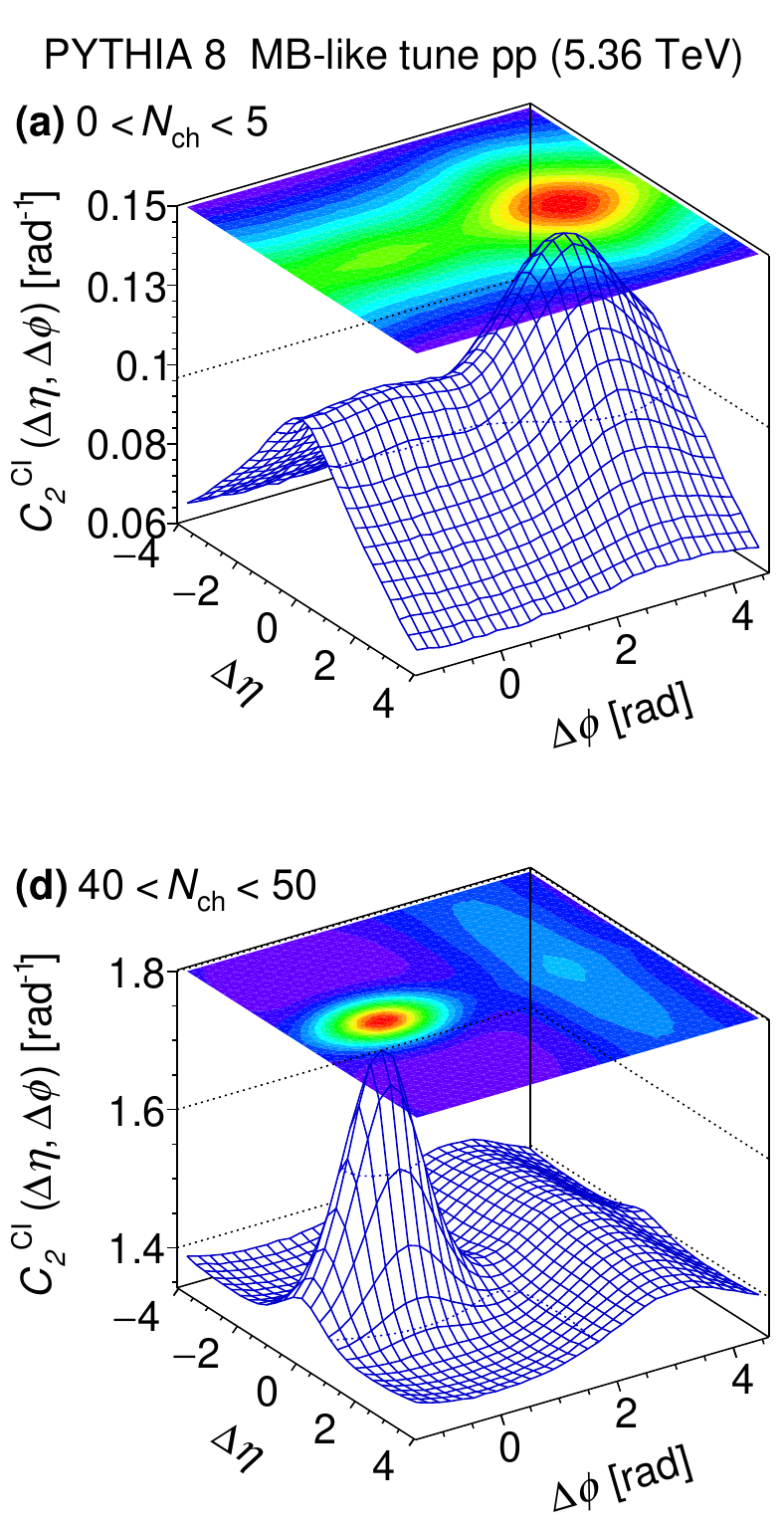}
        \includegraphics[width=0.3\textwidth]{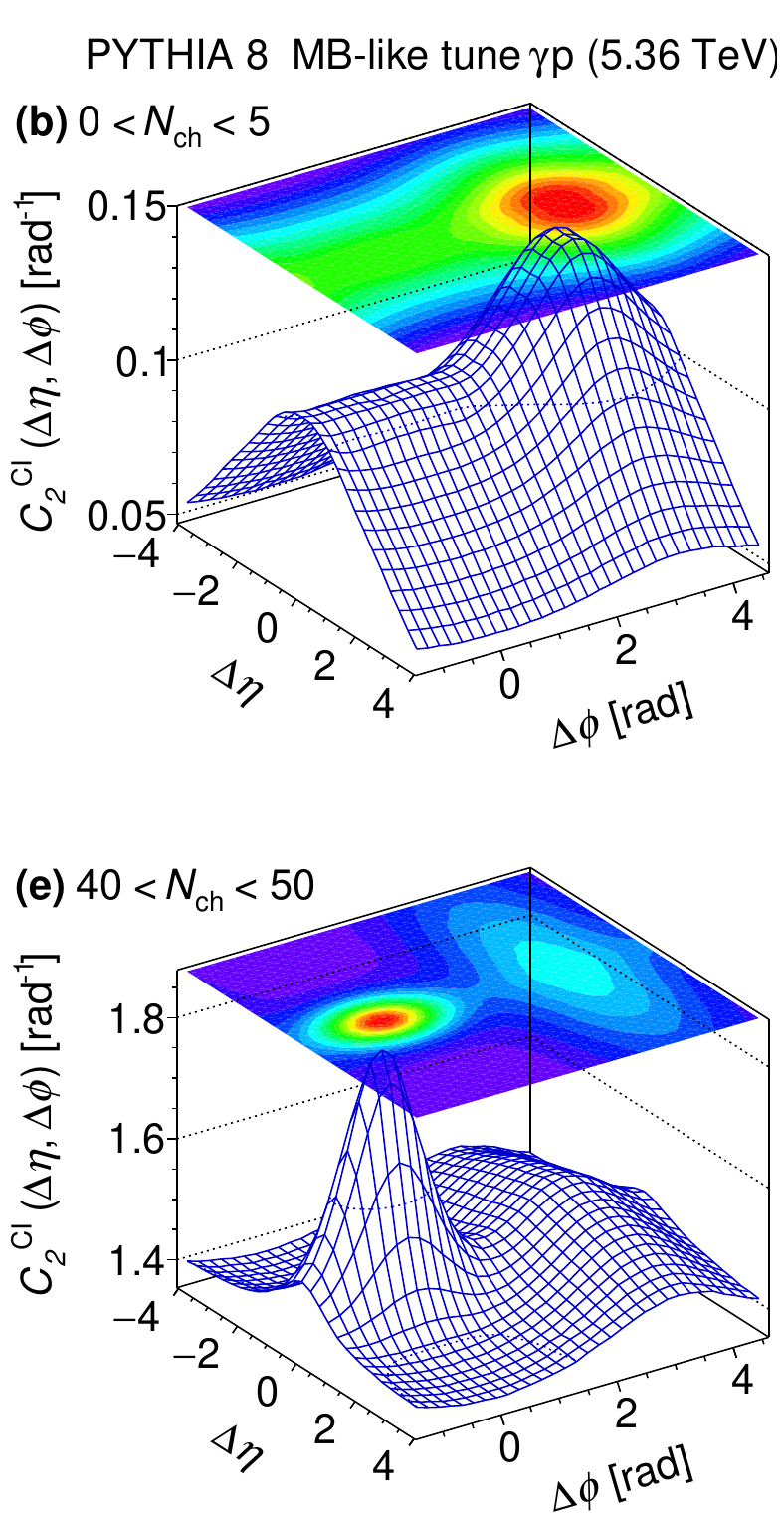 }
        \includegraphics[width=0.3\textwidth]{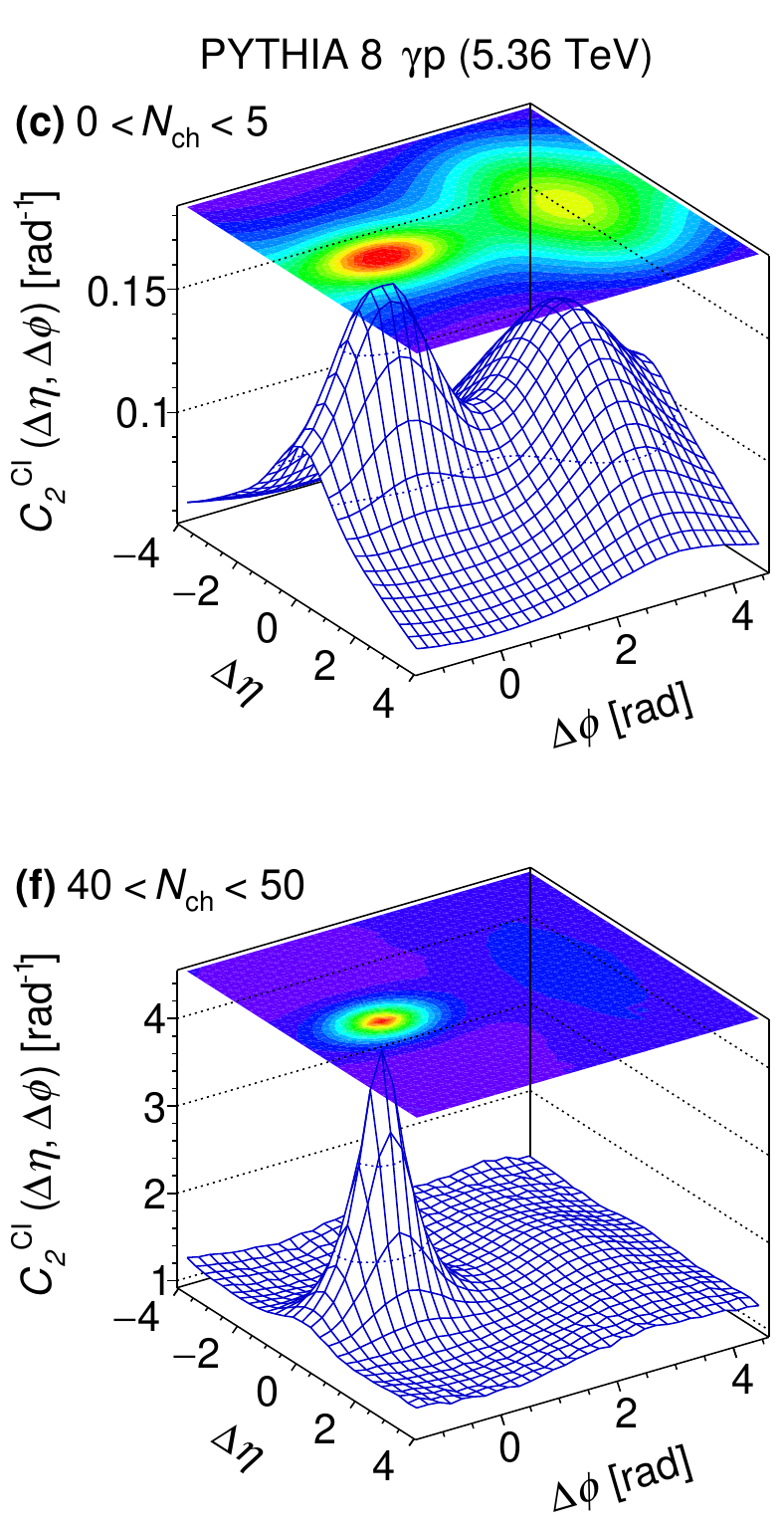}
\caption{Two-dimensional charge-independent number correlations ($C_{2}^\mathrm{CI}$) for $0 < N_{\mathrm{ch}} < 5$ (upper panel) and $40 < N_{\mathrm{ch}} < 50$ (lower panel) in \textsc{mb}-like tune pp (left), \textsc{mb}-like tune \gp (middle), and \gp (right) 
collisions at $\sqrt{s}=5.36$ TeV from \pythia. The results are obtained in the kinematic region $|\eta| < 2.4$ and $0.3 < p_{\mathrm{T}} < 3.0$ GeV charged hadrons.}
\label{fig:comp_CI_corr_c2}
\end{figure*}

This section presents the charge-independent correlations $C_{2}^\mathrm{CI}$ and, $P_{2}^\mathrm{CI}$ for charged hadrons, calculated as the sum of the SS and OS correlations, discussed in Eqs.~(\ref{eqn_balfun_c2ci}) and (\ref{eqn_balfun_p2}). 
Figure~\ref{fig:comp_CI_corr_c2} shows $C_{2}^\mathrm{CI}$ correlations for charged hadrons in
MB pp collisions (left column), \gp collisions with \textsc{mb}-like tune (middle column), and \gp collisions (right column) for  two \nch classes, $0 < N_\mathrm{ch} < 5$ (upper panel) and $40 < N_\mathrm{ch}< 50$ (lower panel). \\

For the low \nch, the near-side CI correlations in MB pp and \textsc{mb}-like \gp exhibit relative suppression with respect to \gp collisions.  This behavior can be attributed to the reduced contribution of short-range correlations and multiparticle
interactions. A clear nearside enhancement emerges in all three configurations at higher \nch. The enhancement with multiplicity is associated with a larger contribution from MPI and semi-hard processes, which strengthen short-range correlations and sharpen the nearside peak; however, the away-side structure becomes relatively suppressed.\\

%%%-------%% ------C2CI-----------------------------

\begin{figure*}  % P2CI correlations
    \centering
        \includegraphics[width=0.3\textwidth]{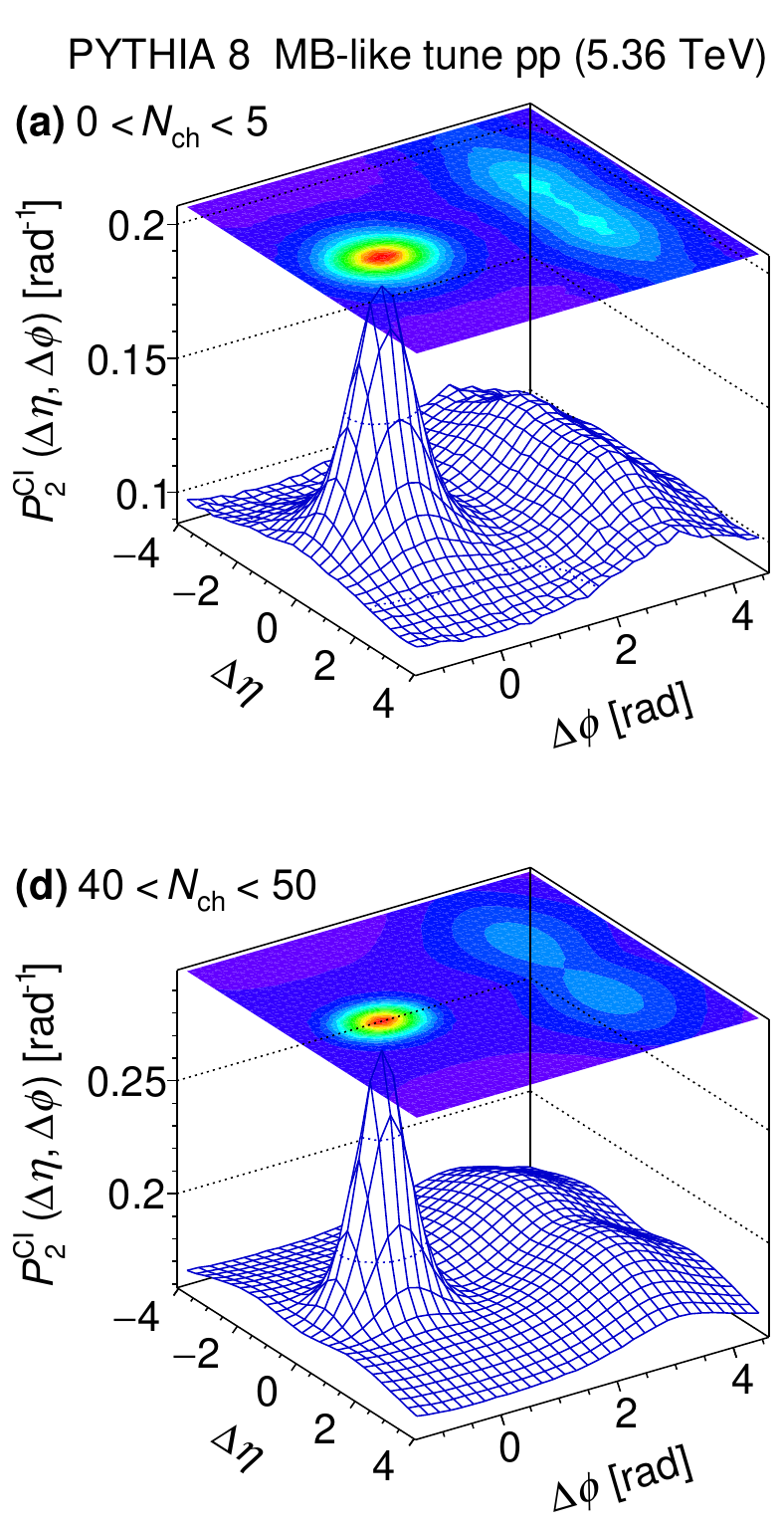}
        \includegraphics[width=0.3\textwidth]{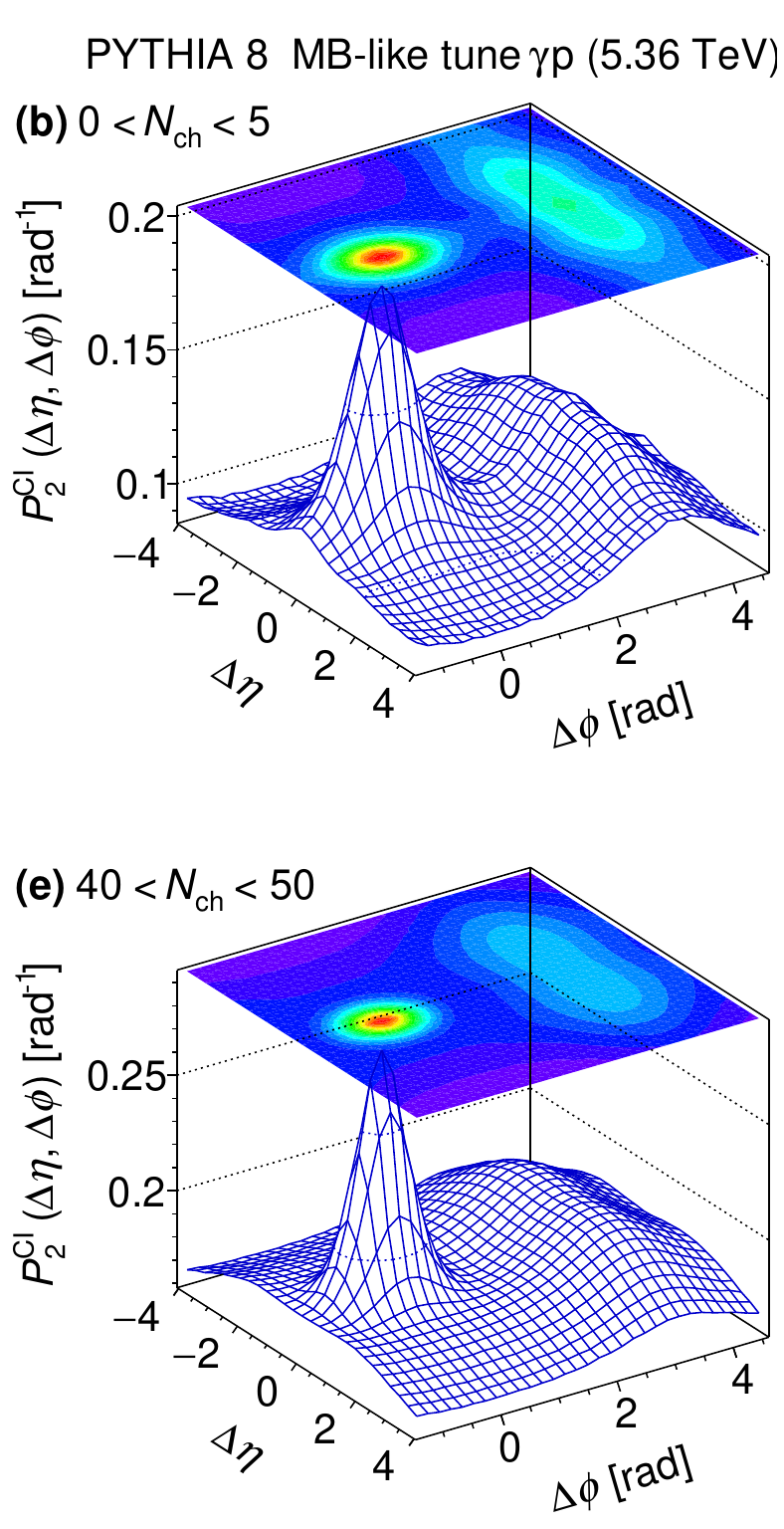}
        \includegraphics[width=0.3\textwidth]{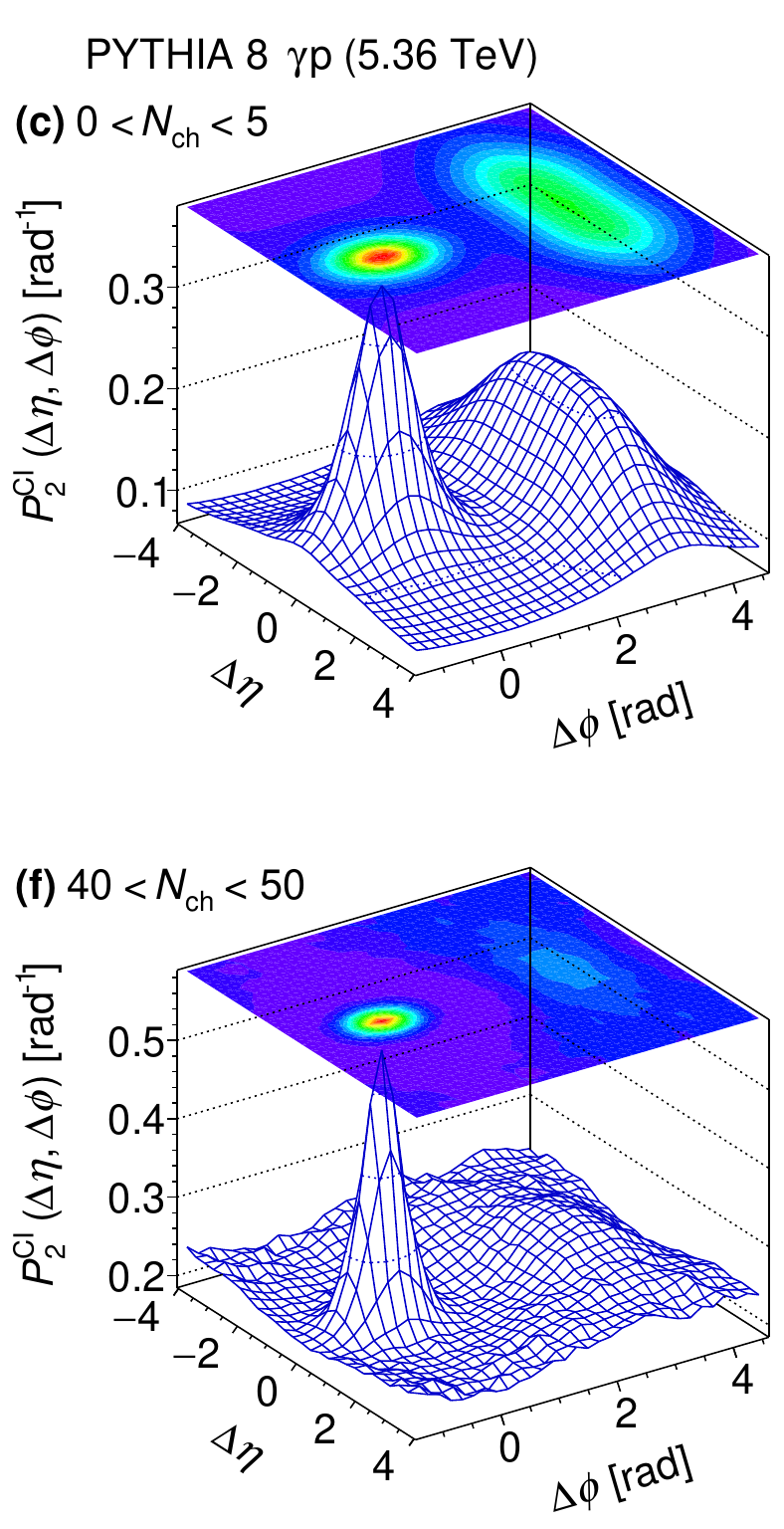}
\caption{Two-dimensional charge-independent momentum correlations ($P_{2}^\mathrm{CI}$) for $0 < N_{\mathrm{ch}} < 5$ and $40 < N_{\mathrm{ch}} < 50$ in \textsc{mb}-like tune pp (left), \textsc{mb}-like tune \gp (middle), and \gp (right) 
collisions at $\sqrt{s}=5.36$ TeV from \pythia. The results are obtained in the kinematic region $|\eta| < 2.4$ and $0.3 < p_{\mathrm{T}} < 3.0$ GeV for charged-hadrons.}
\label{fig:comp_CI_corr_p2}
\end{figure*}

Similarly, Figure~\ref{fig:comp_CI_corr_p2} presents CI correlations constructed from the transverse momentum components \(P_{2}(\Delta\eta,\Delta\phi)\) for two different event configurations. For low-multiplicity events, all these configurations exhibit a localized nearside peak around \((\Delta\eta,\Delta\phi) \approx (0,0)\), indicating short-range correlations dominated by jet fragmentation, resonance decays, and local charge conservation. However, the nearside peak in \gp collisions is visibly stronger and more sharply localized compared to MB pp and \textsc{mb}-like \gp. As the event multiplicity increases, a pronounced enhancement of the nearside peak is observed in both configurations. In contrast, the \gp configuration shows a significantly larger peak magnitude and a reduced away-side structure, indicating that the correlations are dominated by hard scattering processes as the \mpt increases. \\

The CI and CD correlations of $C_{2}$ and $P_{2}$ are studied for charged hadrons in pp and \gp\ collisions. A similar behavior is observed for both CI and CD results, however, the amplitude of the CI correlations is larger than that of the CD correlations because the CI component includes contributions from both SS and OS pairs. This suggests that the final CI correlation strength arises from various physics processes, such as jet fragmentation, resonance decays, the Hanbury Brown-Twiss (HBT) effect, local charge conservation (LCC), and other global event properties. In addition, the nearside peak of $P_{2}^\mathrm{CI}$ is found to be narrower than $C_{2}^\mathrm{CI}$. 
This behavior is consistent with the angular ordering of the \pt, where produced particles exhibit stronger localized momentum correlations. Similar behavior is also expected from hadronic resonance decays as discussed in Refs.~\cite{alicep2r2pp, alicebfpbpbpb, Behera:2025ymi}.

%%%%% ------P2CI-------

%%%%%%%%%%%%%%%%%%%%%%%%%%%%%%%%%%%%%%%%%%%%%%%%%%%

\end{document}